\documentclass[twocolumn]{aastex631}

\def\msun{\hbox{M$_\odot$}}
\newcommand{\HI}{\mbox{H~{\sc i}}}
\newcommand{\HII}{\mbox{H~{\sc ii}}}
\newcommand{\AlII}{\mbox{Al~{\sc ii}}}
\newcommand{\SiIII}{\mbox{Si~{\sc iii}}}
\newcommand{\SiII}{\mbox{Si~{\sc ii}}}
\newcommand{\SII}{\mbox{S~{\sc ii}}}
\newcommand{\CII}{\mbox{C~{\sc ii}}}
\newcommand{\CIII}{\mbox{C~{\sc iii}}}
\newcommand{\SiIV}{\mbox{Si~{\sc iv}}}
\newcommand{\CIV}{\mbox{C~{\sc iv}}}
\newcommand{\NII}{\mbox{N~{\sc ii}}}
\newcommand{\NV}{\mbox{N~{\sc v}}}
\newcommand{\SV}{\mbox{S~{\sc v}}}

\shorttitle{CLUE\SII}
\shortauthors{Sirressi et al.}

\begin{document}

\title{CLusters in the Uv as EngineS (CLUES). II. Sub-kpc scale outflows driven by stellar feedback}

\correspondingauthor{Mattia Sirressi}
\email{mattia.sirre@gmail.com}

\author{Mattia Sirressi}
\affiliation{Department of Astronomy, Oskar Klein Centre, Stockholm University, AlbaNova University
Centre, SE-106 91 Stockholm, Sweden}

\author[0000-0002-0786-7307]{Angela Adamo}
\affiliation{Department of Astronomy, Oskar Klein Centre, Stockholm University, AlbaNova University
Centre, SE-106 91 Stockholm, Sweden}

\author{Matthew Hayes}
\affiliation{Department of Astronomy, Oskar Klein Centre, Stockholm University, AlbaNova University
Centre, SE-106 91 Stockholm, Sweden}

\author{Thøger Emil Rivera-Thorsen}
\affiliation{Department of Astronomy, Oskar Klein Centre, Stockholm University, AlbaNova University
Centre, SE-106 91 Stockholm, Sweden}


\author[0000-0003-4137-882X]{Alessandra Aloisi}
\affiliation{Space Telescope Science Institute, 3700 San Martin Drive, Baltimore, MD 21218, USA}

\author[0000-0001-8068-0891]{Arjan Bik}
\affiliation{Department of Astronomy, Oskar Klein Centre, Stockholm University, AlbaNova University
Centre, SE-106 91 Stockholm, Sweden}

\author[0000-0002-5189-8004]{Daniela Calzetti}
\affiliation{Department of Astronomy, University of Massachusetts, 710 N. Pleasant Street, LGRT 619J, Amherst, MA 01002, USA}

\author{John Chisholm}
\affiliation{Department of Astronomy, The University of Texas at Austin, 2515 Speedway, Stop C1400, Austin, TX 78712, USA}

\author[0000-0003-0724-4115]{Andrew J. Fox}
\affiliation{AURA for ESA, Space Telescope Science Institute, 3700 San Martin Drive, Baltimore, MD 21218, USA}
\affiliation{Department of Physics \& Astronomy, Johns Hopkins University, 3400 N. Charles St., Baltimore, MD 21218}

\author{Michele Fumagalli}
\affiliation{Università degli Studi di Milano-Bicocca, Dip. di Fisica G. Occhialini, Piazza della Scienza 3, 20126 Milano, Italy}
\affiliation{INAF – Osservatorio Astronomico di Trieste, via G. B. Tiepolo 11, I-34143 Trieste, Italy}

\author[0000-0002-3247-5321]{Kathryn Grasha}
\affiliation{Research School of Astronomy and Astrophysics, Australian National University, Canberra, ACT 2611, Australia}   
\affiliation{ARC Centre of Excellence for All Sky Astrophysics in 3 Dimensions (ASTRO 3D), Australia}   
\affiliation{Visiting Fellow, Harvard-Smithsonian Center for Astrophysics, 60 Garden Street, Cambridge, MA 02138, USA}    

\author[0000-0003-4857-8699]{Svea Hernandez}
\affiliation{AURA for ESA, Space Telescope Science Institute, 3700 San Martin Drive, Baltimore, MD 21218, USA}

\author[0000-0003-1427-2456]{Matteo Messa}
\affiliation{Observatoire de Genève, Université de Genève, Chemin Pegasi 51, Versoix CH-1290, Switzerland}
\affiliation{Department of Astronomy, Oskar Klein Centre, Stockholm University, AlbaNova University
Centre, SE-106 91 Stockholm, Sweden}

\author{Shannon Osborne}
\affiliation{Space Telescope Science Institute, 3700 San Martin Drive, Baltimore, MD 21218, USA}

\author{G{\"o}ran {\"O}stlin}
\affiliation{Department of Astronomy, Oskar Klein Centre, Stockholm University, AlbaNova University
Centre, SE-106 91 Stockholm, Sweden}

\author[0000-0003-2954-7643]{Elena Sabbi}
\affiliation{Space Telescope Science Institute, 3700 San Martin Drive, Baltimore, MD 21218, USA}

\author[0000-0002-3933-7677]{Eva Schinnerer}
\affiliation{Max Planck Institute for Astronomy, K{\"o}nigstuhl 17,
69117 Heidelberg, Germany}

\author[0000-0002-0806-168X]{Linda J. Smith}
\affiliation{Space Telescope Science Institute, 3700 San Martin Drive, Baltimore, MD 21218, USA}

\author{Christopher Usher}
\affiliation{Department of Astronomy, Oskar Klein Centre, Stockholm University, AlbaNova University
Centre, SE-106 91 Stockholm, Sweden}

\author[0000-0001-8289-3428]{Aida Wofford}
\affiliation{Instituto de Astronom\'{i}a, Universidad Nacional Aut\'{o}noma de M\'{e}xico, Unidad Acad\'{e}mica en Ensenada, Km 103 Carr. Tijuana-Ensenada, Ensenada 22860, M\'{e}xico}





\begin{abstract}
We analyze the far-ultraviolet (FUV, 1130 $\rm \AA$ to 1770 $\rm \AA$ restframe) spectroscopy of 20 young ($<$50 Myr) and massive ($>10^4$ \msun) star clusters (YSCs) in 11 nearby star-forming galaxies. We probe the interstellar gas intervening along the line of sight, detecting several metal absorption lines of a wide range of ionization potentials, from 6.0 eV to 77.5 eV.
Multiple-component Voigt fits to the absorption lines are used to study the kinematics of the gas. We find that nearly all targets in the sample feature gas outflowing from 30 km~s$^{-1}$ up to 190 km~s$^{-1}$, often both in the neutral and ionized phase. The outflow velocities correlate with the underlying stellar population properties directly linked to the feedback: the mass of the YSCs, the photon production rate and the instantaneous mechanical luminosity produced by stellar winds and SNe. We detect a neutral inflow in 4 targets, which we interpret as likely not associated with the star cluster but tracing larger scale gas kinematics. A comparison between the outflows’ energy and that produced by the associated young stellar populations suggests an average coupling efficiency of 10 \% with a broad scatter. 
Our results extend the relation found in previous works between galactic outflows and the host galaxy star-formation rate to smaller scales, pointing towards the key role that clustered star formation and feedback play in regulating galaxy growth.


\end{abstract}



\section{Introduction} \label{sec:intro}

Star formation and related feedback play a major role in shaping the evolution of galaxies.
Large amounts of energy, momentum and ionizing radiation produced by stellar feedback \citep[e.g.]{krumholz2014, Krumholz2019} have an important impact on the physical conditions of the interstellar medium (ISM) of the host galaxy. 
Multi-phase outflows are typically observed in star-forming galaxies \citep[for a review see][]{Veilleux2020}. These galactic-scale outflows are fuelled by compact star-forming regions and massive star clusters \citep[e.g.][]{smith2006, westmoquette2013}, which are the sources of the energy and momentum driving the outflows. Indeed, several studies have shown that simulations of galaxies that do not implement clustered feedback in their physics, result in an unrealistic ISM and form new stars at an unrealistically high rate \citep[e.g.][]{Shetty2012, Kim2017, smith2020}
 
Cosmological simulations can hardly resolve the small spatial scales of feedback processes and at the same time reproduce galactic scale phenomena such as kpc-scale outflows. As a consequence, different recipes are needed in these simulations to incorporate feedback as a function of other parameters \citep[e.g. the star formation rate $SFR$ or the stellar mass $M\star$,][]{Somerville2015,Naab2017}. The parametrized scaling relations used for this purpose are calibrated by reference to observations \citep[e.g.][]{Vogelsberger2013, Crain2015, Pillepich2018}. Informing this type of simulations with accurate stellar properties and outflow measurements is a major motivation for our work.

To study the effect that clustered feedback has on the natal cloud as well as the immediate ISM, isolated clouds or clouds surrounded by an idealized disk plane medium are typically simulated \citep[e.g.][]{Lancaster2021, dobbs2022, Bending2022, grudic2022}. These simulations find that cluster feedback drives gas outflows, also referred to as bubbles and superbubbles. A complex problem to tackle in these studies is the interaction of the stellar feedback energy and momentum with the surrounding medium. \cite{Lancaster2021} account for a turbulent and fractal ISM around the star cluster rather than the typically assumed uniform ambient medium, which is likely far from reality. They found that the large interface area offered by the fractal ISM favors strong cooling and makes the outflow momentum-driven rather than energy-driven.

One other important property of the ISM gas morphology around young star clusters is the presence of channels through which the feedback energy and momentum can escape, especially when pre-SN feedback, i.e. stellar winds, has already occurred and opened those channels \citep{Lucas2020}. Moreover, given the multi-phase nature of the ISM, different phases of the gas might have a different coupling efficiency with the feedback energy \citep{Haid2018}, i.e. they will absorb the feedback momentum and be accelerated to different degrees. \cite{Kim2020} presented pc-resolution simulations of a galactic disk in a project named Simulating Multiscale Astrophysics to Understand Galaxies (SMAUG). They stressed the importance of including at least two distinct gas components in cosmological simulations: the mass-delivering cool gas ($T \sim 10^4$ K) and the energy/metal-delivering hot gas ($T \gtrsim 10^6$ K). Observations are key to understand better the impact of stellar feedback on the ISM and the driving mechanisms of outflows. 

In this study, we primarily make use of FUV observations provided by the Hubble Space Telescope (HST) to trace the neutral and ionized phases of the atomic gas.
FUV spectroscopy contains a wealth of information, not only about the stellar population \citep{Wofford2013,Hernandez2019,chisholm2019,Leitherer2020,Sirressi2022,Hayes2023} but also about the ISM that lies along the line of sight to the starburst. The gas absorbs the FUV light produced by the FUV continuum source, allowing the observer to infer the kinematics, the column density and the size of the gas clouds in the surrounding ISM, and tracing for example outflowing gas. Several absorption lines have been detected in the spectral FUV window, which probe a wide range of ionization potentials \citep[from 6.0 eV for \AlII$\,$ 1671 Å to 77.5 eV for \NV$\,$ 1238,1242 Å][]{Shapley2003,Grimes2009,Steidel2010,Kornei2012,Heckman2015,Chisholm2015,Chisholm2017,chisholm2018,Sugahara2019,Ostlin2021,Xu2022,Hayes2023}.\footnote{Hereafter we will omit the symbol $\lambda$ in the nomenclature of the spectral lines.}
Many of these works have been carried out using the same spectrograph adopted here, the Cosmic Origins Spectrograph (COS), but with galaxies at an average redshift of $z \sim 0.1$ 
probing large portions or the entire galaxies (kpc scales) and their galactic scale outflows. 
For the first time to our knowledge, in this study we employ the same technique using FUV data to investigate the properties of stellar feedback and of the ISM gas at tens of parsec scales, zooming into the regions where the feedback originates. 


A number of relations have been found in medium-size samples (with several tens) of star-forming and starburst galaxies, such as the increase of galactic outflow velocities as a function of SFR, SFR per unit area and SFR per unit mass \citep{Heckman2015,Chisholm2015,Xu2022}. This confirms the link between star formation and ISM gas, and more specifically the role of stellar feedback as the driver of galactic outflows. In particular, \cite{Xu2022} measured the outflow energetics such as the rates at which mass, energy and momentum are being transported out of the center of the galaxy, probing spatial scales of 0.1-10.7 kpc. They found that these quantities correlate with the corresponding rates of the feedback energetics.
Observational studies at small scales and in a sample of more typical galaxies, like the one presented here, are also crucial for testing theoretical models and numerical simulations that attempt to implement feedback and spatially resolve outflows without the aid of the recipes mentioned above \citep{Schneider2018,Schneider2020}. 

In this work, we use FUV spectroscopy to measure key feedback quantities of a sample of 20 YSCs and investigate their effects on the surrounding ISM. The survey, named CLusters in the Uv as EngineS (CLUES) is presented in a previous paper \citep[][hereafter Paper I]{Sirressi2022}. CLUES is a COS campaign aimed at acquiring the 1130 to 1770 $\rm \AA$ rest-frame spectroscopy of 20 very young ($<50$ Myr) and massive ($> 10^4\, \msun$) star clusters. The targets of this study are drawn from the Local Volume galaxy population previously observed with the Hubble treasury program Legacy ExtraGalactic Uv Survey \citep[LEGUS,][]{Calzetti2015}. We have selected in most cases two clusters per galaxy, at various distances from the center of the galaxy (from 0.1 to 2.8 half-mass radii). The ultimate goal of this project is to connect the kinematics of the outflowing gas with the stellar feedback injected by individual young star clusters, which has not been fully accomplished from previous observational studies. 
In Paper I (see Table 3), we have derived the age, mass, metallicity and attenuation of the clusters from both multi-wavelength photometry and FUV spectroscopy with the HST. 
In the present study we quantify the ionizing radiation in terms of ionizing photons per second, as well as the mechanical energy and mechanical luminosity produced by the stars using the stellar population synthesis models \citep[based on Starburst99, ]{leitherer1999}{} reported in Paper I. In addition, based on the same FUV data, we analyze here the properties of the intervening ISM: kinematics, column density, ionization state, outflow mass and outflow energetics.


We present the observations and the data in Section \ref{sec:data}. In Section \ref{sec:feedback} we report the stellar feedback quantities derived from the stellar population models described in Paper I. In Section \ref{sec:kinematics} we present the kinematics of the ISM gas modelled with multiple Voigt components. In Section \ref{sec:outflows} we show the correlation between outflow velocities and stellar properties. In addition, we estimate the mass and energetics of the detected outflows. We discuss and interpret our results in Section \ref{sec:discussion} and in Section \ref{sec:conclusions} we summarize our work.

\section{Data} \label{sec:data}

\subsection{Survey}
CLUES is an HST medium-size program (ID 15627, PI: Adamo) that provides FUV spectroscopy of 20 YSCs in local galaxies (less than 16 Mpc away). The data were taken with the COS spectrograph using two gratings: G130M and G160M. Three sources (NGC4449-YSC1, NGC5253-YSC1 and NGC5253-YSC2) were observed as part of an earlier program (ID 11579, PI: Aloisi). All our data products are available at MAST as a High Level Science Product via \dataset[10.17909/2bay-1d16]{\doi{10.17909/2bay-1d16}}. The spectra have an average S/N of about 40 when resampled to 0.4 Å, which is the resolution of the stellar population synthesis models. We refer to Paper I for the observational setup and for more details about the data.

The CLUES clusters have ages between 1 and 50 Myr old, masses in the range $10^4$ - $10^6\, \rm M_{\odot}$. The host galaxies cover a wide range of star formation activity (SFR = 0.1-6.8 $\rm M_{\odot} yr^{-1}$), metallicity (12+log(O/H)=8.0-9.0) and morphology (grand-design arms, circumnuclear starburst rings, flocculent spirals, dwarf starburst and tidal features).
For a detailed description of the sample selection, data acquisition and main properties of the target galaxies, see Section 2 and 3 of Paper I.

\begin{figure*}
    \centering
    \includegraphics[width=\textwidth]{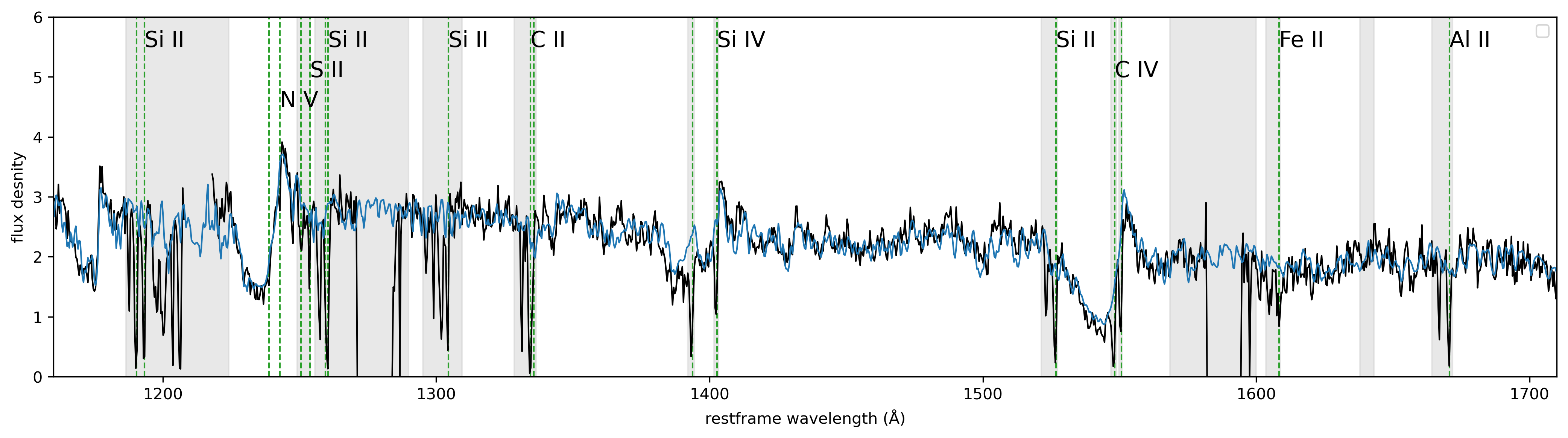}
     \caption{Observed FUV spectrum of M74-YSC1 in black, best-fit model in blue. The regions masked out of the fit are marked in gray. The detected ISM absorption lines are marked as green dashed lines and labeled. See Table \ref{tab:ISM_lines} for the full list of detected absorption lines.}
    \label{fig:M74-1_spectrum_lines}
\end{figure*}

Figure \ref{fig:M74-1_spectrum_lines} shows the FUV spectrum of the target M74-YSC1 as an example, illustrating the detected ISM absorption lines, which allow us to study the gas kinematics.

\subsection{Spectral resolution \& foreground contamination}
 A precise measurement of the spectroscopic resolving power is crucial to study the kinematics of the gas. Therefore, we derive the effective spectral resolution of our observations by convolving the profile of the extended UV source (in the acquisition image) with the Line Spread Function (LSF) of the COS instrument. Table \ref{tab:res} reports the values of the effective spectral resolution for each target in km~s$^{-1}$. The measured FWHMs are a few tens of km~s$^{-1}$, typically lower than the velocities that we measure for the star clusters hosting an outflow or an inflow. We note that the median resolution of our targets is 38 km~s$^{-1}$, which corresponds to 3 resolution elements (18 COS FUV pixels). For comparison, the resolution of a point source observed with COS at lifetime position 4 (LP4) as the CLUES targets, at the wavelengths of our data, is 15-25 km~s$^{-1}$, whereas the resolution of a fully extended source (i.e. with the size of the COS aperture), would be 190-194 km~s$^{-1}$.

\begin{table}
    \centering
    \begin{tabular}{lr}
    \hline
    source name & res. (km~s$^{-1}$) \\   
\hline
 M51-YSC1     &  34 \\
 M51-YSC2     &  32 \\
 M95-YSC1     &  32 \\
 NGC1313-YSC1 &  38 \\
 NGC1566-YSC1 &  66 \\
 NGC1566-YSC2 &  36 \\
 NGC4485-YSC1 &  51 \\
 NGC4656-YSC2 &  23 \\
 NGC7793-YSC1 &  30 \\
 M74-YSC2     &  34 \\
 NGC1313-YSC2 &  38 \\
 NGC1512-YSC2 &  30 \\
 NGC4656-YSC1 &  28 \\
 NGC7793-YSC2 &  64 \\
 NGC4485-YSC2 &  32 \\
 NGC1512-YSC1 &  26 \\
 M74-YSC1     &  43 \\
 NGC4449-YSC1 &  90 \\
 NGC5253-YSC1 &  47 \\
 NGC5253-YSC2 &  47 \\
\hline
    \end{tabular}
    \caption{Effective spectral resolution of the CLUES observations in km~s$^{-1}$. The values reported have been computed as the FWHM of the convolution between the profile of the extended source and the COS LSF.}
    \label{tab:res}
\end{table}

\begin{figure*}
    \centering
    \includegraphics[width=\textwidth]{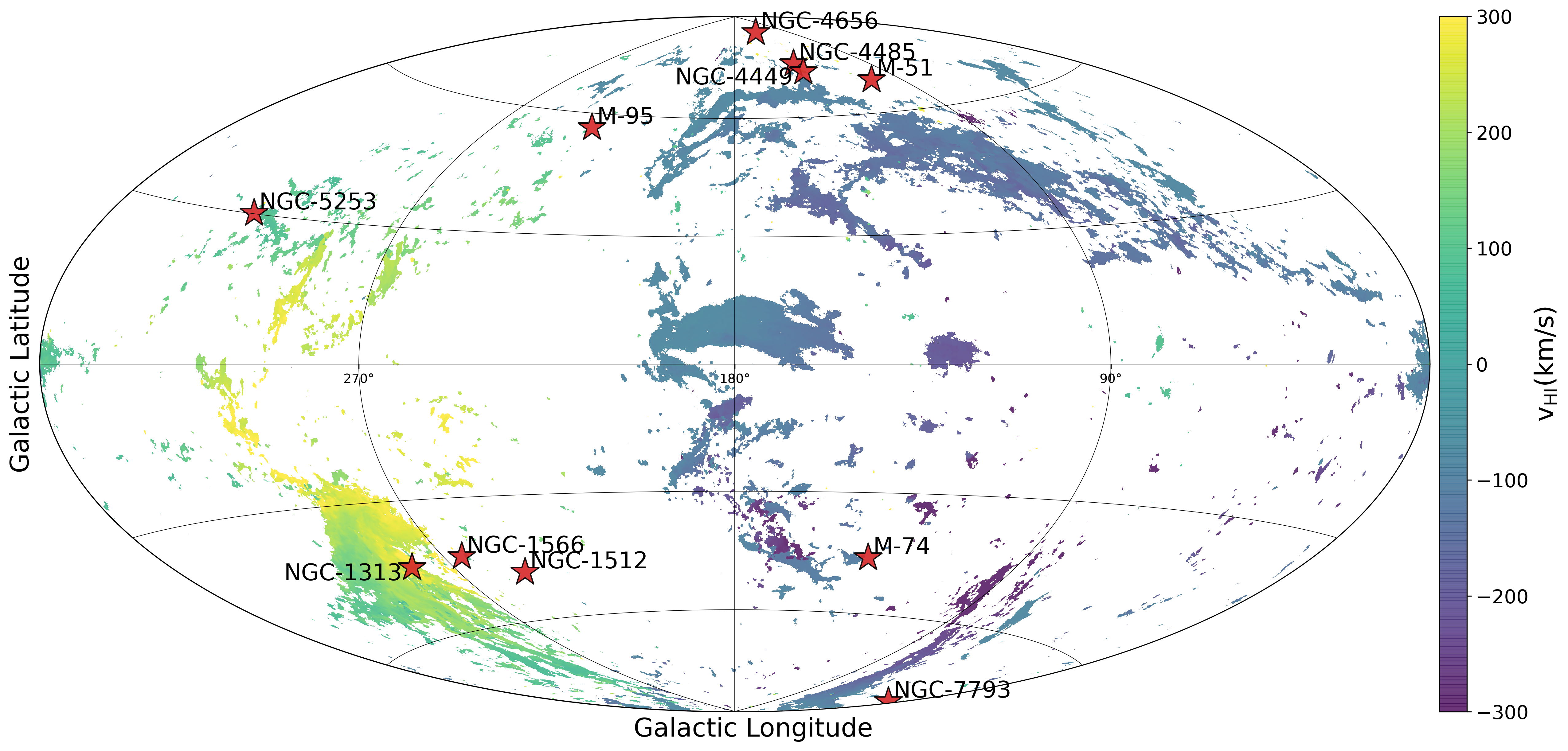}
     \caption{Full-sky map of Magellanic and Galactic high-velocity clouds, color-coded by the LSR velocity of \HI$\,$ 21 cm emission. The locations of the CLUES target galaxies are marked with red stars. The data for the velocity map are taken from \cite{Westmeier2018}}
    \label{fig:sky}
\end{figure*}

Of the 20 targets presented in Paper I, two  (NGC1512-YSC1, NGC4449-YSC1) have been excluded from this analysis. The data of NGC1512-YSC1 have a low signal-to-noise ratio (see Table 1 in Paper I), which is not sufficient for extracting reliable information. The ISM lines of NGC4449-YSC1 suffer from severe contamination from foreground gas along the line of sight not connected with the system.
Due to the low redshift of our targets, another potential issue is, in fact, the blending of the galaxy's ISM absorption lines with foreground absorption lines from the Milky Way or the Magellanic Clouds. This is the case especially for the data of NGC4449-YSC1, the target with the lowest recession velocity in our sample, that show a broad component, making it challenging to distinguish the components of the outflowing gas. Figure \ref{fig:sky} illustrates the locations of the targeted galaxies in the sky, plotted on top of a velocity map of the gas in the Magellanic System and Milky Way. The data of NGC7793-YSC1, NGC7793-YSC2, NGC1313-YSC1, NGC1313-YSC1, NGC5253-YSC1, NGC5253-YSC1 show the signature of gas absorption at a velocity between 200 and 400 km~s$^{-1}$ caused by the Magellanic system, which lies along the line of sight. We keep these sources in our sample as we are able to distinguish the contaminating absorption components and to remove them. 

\subsection{Other data used in this analysis}
The FUV data feature a few stellar absorption lines that can be used to derive the redshift of the targets, e.g. \CIII$\,$1247 and \CIII$\,$1176. However, this is challenging as the observed absorption is the blending of multiple transitions at different wavelengths. For this reason, we also made use of integral field spectrograph (IFS) MUSE data of the clusters that have archival observations (8 out of 20 targets) to extract the profile of the H$\alpha$ emission line and derive the redshift of the target star cluster (see Section \ref{sec:redshifts}). The MUSE-PHANGS sample \citep{Emsellem2022} covers the following CLUES YSCs: M95-YSC1, NGC1566-YSC1, NGC1566-YSC2, M74-YSC2 and NGC1512-YSC2. The MUSE data of NGC1313-YSC2 are instead taken from the program 106.21Q6 (PI L. Della Bruna), whereas the MUSE data of NGC7793-YSC1 and NGC7793-YSC2 from the program 097.B-0899 (PI E. Ibar). We extracted 5 arcsec side cubes centered on the clusters from the larger cubes to facilitate the analysis. One dimensional spectrum is extracted using a circular aperture of equal size to the COS aperture. No further corrections are applied, since the only measurement made consists of the centroid position of the H$\alpha$ emission line.

\subsection{Comparison samples}
We report here the details of a few samples of galactic outflows from the literature that we will compare to our sample in Section \ref{sec:discussion}.
\cite{Chisholm2015} reported on a sample of 48 nearby star-forming galaxies hosting galactic outflows detected using a combination of different \SiII$\,$ transitions observed with the COS spectrograph. In 15 out of 48 sources in the COS spectra are targeting individual star clusters similarly to our work. We note that in \cite{Chisholm2015} the following targets are in common with our work: NGC5253-1, NGC5253-2. 
\cite{Heckman2015} investigated a sample of 39 low-redshift galaxies hosting starburst-driven galactic outflows. We note that the works of \cite{Heckman2015} and \cite{Chisholm2015} have 10 targets in common.
\cite{Heckman2015} used the strongest absorption lines for deriving the outflow velocities: \SiIII$\,$1206 for the COS data, \CIII$\,$977 and \NII$\,$1084 for the FUSE data.
Half the sample has large outflow velocities (300-500 km~s$^{-1}$), the other half has lower outflow velocities (50-200 km~s$^{-1}$). We note that neither \cite{Heckman2015} nor \cite{Chisholm2015} remove the zero-velocity component to disentangle the stationary gas, like in our work (Sec \ref{sec:voigt}). Therefore, their average velocities might be underestimated.
In both works, the SFR is measured using the relation in \cite{kennicutt2012} based on the UV and IR luminosity. In addition, they measured the specific SFR (SFR per stellar mass) and the $\Sigma_{SFR}$ (SFR per unit area) of the galaxy. 
In \cite{Chisholm2015}, $\Sigma_{SFR}$ is computed for the area of the COS aperture - e.g. using the SFR in the aperture and the area of the aperture, similarly to this study (Sec \ref{sec:literature_comparison}). Whereas \cite{Heckman2015} measured the UV half-light radius and assumed that half the SFR is within it.

\cite{Xu2022} derived the properties of starburst-driven warm ionized outflows for a sample of 45 low redshift galaxies named CLASSY (COS Legacy Archive Spectroscopic SurveY) as well as five additional targets with COS archival data. 
We note that many of the CLASSY dwarfs are galaxies that contain only one bright star cluster.  They measured the outflow kinematics with a similar method to the one applied in this work but using Gaussian rather than Voigt functions. They did not tie the velocity structure of all transitions, but they computed an average of the values found for each detected absorption, considering both low and high ionization potentials. They derived the stellar mass and the SFR from the spectral energy distribution (SED) fitting to the Galaxy Evolution Explorer and Sloan Digital Sky Survey (SDSS) data, UV+optical photometry \citep[for more details see]{Berg2022}. However, in this case as well as in \cite{Chisholm2015} and \cite{Heckman2015}, the majority of the galaxies are at large distances ($>$ 80 Mpc) and the stellar physical properties (e.g. SFR) are determined for the whole galaxy.

\section{Stellar feedback} \label{sec:feedback}

In Paper I, we presented various models for the FUV spectroscopy of the CLUES clusters and showed that in most cases (16 out of 20) two model populations are necessary to describe the stellar content within the COS aperture, whereas in the remaining 4 cases only one population is sufficient to reproduce the stellar emission. The derived physical values (age, metallicity, mass) are in agreement with those derived with an alternative method described in \cite{chisholm2019} that weighs the contribution of several single stellar populations, confirming that our determinations are quite robust. The reader can find more details on the methods mentioned above in \cite{chisholm2019} and Paper I. 

We took our fiducial stellar population models and the corresponding best-fit parameters and derived feedback quantities. We did this by using the stellar feedback quantities tabulated in Starburst 99 libraries using the same model assumptions as in Paper I. 
Finally, we employed the same kind of spline interpolator developed for the stellar population fitting parameters in order to derive continuous values for each feedback quantity. Using the best-fitted physical parameters of the stellar populations (age, mass, extinction, metallicity), we estimated the production rate of photons that can ionize Hydrogen $(Q_H)$ as well as the mechanical energy and mechanical luminosity of two different sources of feedback: stellar winds and supernovae. 

The mechanical luminosity is defined as the mechanical energy produced per unit time. In particular, it is the core-collapse type II SNe that are relevant in this context, and hereafter we will refer to this class just as SNe. We report in Table \ref{tab:feedback_YO} all the above quantities separately for the young (top table) and the older (bottom table) stellar populations of each target modelled with two stellar populations. These two stellar populations do not necessarily coincide with the ``FUV dominant" and ``FUV fainter" population discussed in Paper I, however, we chose to have here the distinction young/old as more relevant in the context of feedback. The feedback quantities for the total stellar population is simply given by the sum of the contributions of the two stellar populations. Because, the age errors returned by our Monte Carlo Simulation are underestimated, we impose a floor of 0.5 Myr on the errors of the age values, equal to half a step of the model grid.

\begin{table*}
\centering
\fontsize{8}{10}\selectfont
\begin{tabular}{lcccccccccc}
\hline
\multicolumn{11}{c}{Young stellar population}\\
\hline
 source name   &        $age $  & $M$ & $f$ & log$(Q_H)$ &  log($E_{winds}$)  &   log($E_{SN}$)  & log($E_{\rm Tot}$) & log($L_{winds}$) &   log($L_{SN}$) &  log($L_{\rm Tot}$)\\
 &    Myr   & $10^6 \msun$ & & photons/s & erg &  erg & erg & erg/s  & erg/s &  erg/s \\
 (1) & (2) & (3) & (4) & (5) & (6) & (7) & (8) & (9) & (10) & (11)\\
\hline
 M74-2     & 2.3$^{+0.5}_{-0.5}$  & 0.04$^{+0.00}_{-0.00}$ & 1.0 & 50.9 & 52.8 &   -   & 52.8 & 39.1 &   -   & 39.1 \\
 M95-1     & 2.5$^{+0.5}_{-0.5}$  & 0.72$^{+0.31}_{-0.00}$ & 0.9 & 52.1 & 54.2 &   -   & 54.2 & 40.3 &   -   & 40.3 \\
 M74-1     & 2.6$^{+0.5}_{-0.5}$  & 0.07$^{+0.03}_{-0.00}$ & 0.9 & 51.1 & 53.1 &   -   & 53.1 & 39.2 &   -   & 39.2 \\
 NGC1512-2 & 2.6$^{+0.5}_{-0.5}$  & 0.03$^{+0.01}_{-0.00}$ & 0.9 & 50.6 & 52.8 &   -   & 52.8 & 38.9 &   -   & 38.9 \\
 M51-1     & 3.4$^{+0.5}_{-0.5}$  & 0.07$^{+0.05}_{-0.00}$ & 0.8 & 50.8 & 53.3 &   -   & 53.3 & 39.1 &   -   & 39.1 \\
 NGC1566-2 & 3.4$^{+4.8}_{-0.5}$  & 0.25$^{+0.63}_{-0.00}$ & 1.0 & 51.4 & 53.8 &   -   & 53.8 & 39.7 &   -   & 39.7 \\
 NGC1512-1 & 3.5$^{+0.5}_{-0.5}$  & 0.03$^{+0.00}_{-0.00}$ & 1.0 & 50.4 & 52.9 &   -   & 52.9 & 38.7 &   -   & 38.7 \\
 NGC7793-1 & 3.5$^{+0.5}_{-0.5}$  & 0.02$^{+0.00}_{-0.00}$ & 1.0 & 50.3 & 52.5 &   -   & 52.5 & 38.5 &   -   & 38.5 \\
 NGC4485-2 & 2.1$^{+0.8}_{-0.5}$  & 0.30$^{+0.05}_{-0.10}$ & 0.3 & 52.2 & 52.9 &   -   & 53.0 & 39.4 &   -   & 39.4 \\
 NGC4485-1 & 4.4$^{+0.5}_{-0.5}$  & 0.05$^{+0.00}_{-0.02}$ & 0.8 & 50.7 & 52.8 &  52.4 & 53.0 & 38.6 &  38.9 & 39.1 \\
 NGC1313-1 & 4.9$^{+0.5}_{-0.5}$  & 0.15$^{+0.03}_{-0.02}$ & 0.9 & 51.1 & 53.4 &  53.1 & 53.5 & 38.9 &  39.4 & 39.6 \\
 NGC1566-1 & 8.0$^{+27.8}_{-0.5}$ & 0.80$^{+0.93}_{-0.00}$ & 1.0 & 51.0 & 54.1 &  54.3 & 54.5 & 38.5 &  40.1 & 40.1 \\
 NGC7793-2 & 9.9$^{+9.4}_{-0.5}$  & 0.01$^{+0.02}_{-0.00}$ & 0.6 & 48.6 & 52.1 &  52.5 & 52.6 & 36.2 &  38.1 & 38.1 \\
 NGC1313-2 & 29.8$^{+17.5}_{-6.9}$& 0.06$^{+0.20}_{-0.00}$ & 0.6 & 47.2 & 52.9 &  53.7 & 53.8 & 34.7 &  38.7 & 38.7 \\
 NGC4656-1 & 4.7$^{+20.7}_{-1.3}$ & 4.59$^{+2.21}_{-1.58}$ & 0.1 & 52.6 & 54.9 &  54.5 & 55.0 & 40.5 &  40.9 & 41.1 \\
 M51-2     & 4.4$^{+32.0}_{-1.2}$ & 0.00$^{+2.35}_{-0.00}$ & 0.4 & 49.1 & 51.5 &  51.0 & 51.6 & 37.3 &  37.4 & 37.7 \\
 NGC4656-2 & 3.1$^{+0.5}_{-0.5}$  & 2.53$^{+0.31}_{-0.32}$ & 0.4 & 52.9 & 54.2 &   -   & 54.2 & 40.8 &   -   & 40.8 \\
 NGC4449-1 & 3.0$^{+0.5}_{-0.5}$  & 0.48$^{+0.09}_{-0.10}$ & 0.4 & 52.2 & 53.2 &   -   & 53.2 & 39.5 &   -   & 39.5 \\
 NGC5253-2 & 3.1$^{+0.5}_{-0.5}$  & 0.04$^{+0.03}_{-0.00}$ & 0.4 & 51.1 & 52.1 &   -   & 52.1 & 38.3 &   -   & 38.3 \\
 NGC5253-1 & 3.0$^{+0.5}_{-0.5}$  & 0.85$^{+0.10}_{-0.20}$ & 0.3 & 52.5 & 53.6 &   -   & 53.6 & 39.9 &   -   & 39.9 \\
\hline
\multicolumn{11}{c}{Old stellar population}\\
\hline
 source name   &        $age $  & $M$ & $f$ & log$(Q_H)$ &  log($E_{winds}$)  &   log($E_{SN}$)  & log($E_{\rm Tot}$) & log($L_{winds}$) &   log($L_{SN}$) &  log($L_{\rm Tot}$)\\
 &    Myr  & $10^6 \msun$ &  & photons/s & erg &  erg & erg & erg/s  & erg/s &  erg/s \\
 (1) & (2) & (3) & (4) & (5) & (6) & (7) & (8) & (9) & (10) & (11)\\
\hline
 M74-2     & - & - & -   & -   & -  & -  & -   & -   & -   & -   \\
 M95-1     & 42.0$^{+5.5}_{-40.1}$  & 6.80$^{+0.00}_{-6.48}$  & 0.1 &   47.4 &   55.4 &   55.9 &   56.0 &   35.8 &   - &   35.8 \\
 M74-1     & 47.4$^{+1.9}_{-44.8}$  & 3.37$^{+2.45}_{-3.34}$  & 0.1 &   47.5 &   55.0 &   55.6 &   55.7 &   34.8 &   - &   34.8 \\
 NGC1512-2 & 47.4$^{+2.1}_{-43.3}$  & 0.10$^{+0.65}_{-0.10}$  & 0.1 &   45.9 &   53.5 &   54.1 &   54.2 &   33.3 &   - &   33.3 \\
 M51-1     & 35.7$^{+7.8}_{-16.5}$  & 0.50$^{+0.10}_{-0.47}$  & 0.2 &   46.7 &   54.3 &   54.8 &   54.9 &   35.0 &   - &   35.0 \\
 NGC1566-2 & - & - & -   & -   & -  & - & -   & -   & -   & -   \\
 NGC1512-1 & 47.4$^{+1.2}_{-41.4}$  & 0.02$^{+0.04}_{-0.02}$  & -  &   45.3 &   52.7 &   53.4 &   53.5 &   32.7 &   - &   32.7 \\
 NGC7793-1 & - & - & -   & -   & -  & - & -   & -   & -   & -   \\
 NGC4485-2 & 3.7$^{+0.0}_{-0.1}$    & 0.01$^{+0.01}_{-0.00}$  & 0.7 &   50.2 &   51.8 &   51.0 &   51.9 &   38.2 &   38.1 &   38.5 \\
 NGC4485-1 & 14.1$^{+30.5}_{-12.1}$ & 0.01$^{+0.72}_{-0.00}$  & 0.2 &   48.2 &   51.9 &   52.8 &   52.9 &   35.2 &   38.2 &   38.2 \\
 NGC1313-1 & 36.1$^{+10.4}_{-30.9}$ & 0.17$^{+0.65}_{-0.15}$  & 0.1 &   46.2 &   53.8 &   54.3 &   54.4 &   34.5 &   - &   34.5 \\
 NGC1566-1 & - & - & -   & -   & -  & - & -   & -   & -   & -   \\
 NGC7793-2 & 47.3$^{+0.0}_{-11.0}$  & 0.24$^{+0.02}_{-0.08}$  & 0.4 &   46.3 &   53.8 &   54.5 &   54.5 &   33.7 &   - &   33.7 \\
 NGC1313-2 & 36.1$^{+11.0}_{-34.2}$ & 1.47$^{+0.88}_{-0.71}$  & 0.4 &   47.2 &   54.7 &   55.3 &   55.4 &   35.5 &   - &   35.5 \\
 NGC4656-1 & 30.1$^{+4.8}_{-0.0}$   & 0.70$^{+0.18}_{-0.26}$ & 0.9 &   48.5 &   53.6 &   54.8 &   54.9 &   35.5 &   39.8 &   39.8 \\
 M51-2     & 36.0$^{+0.4}_{-0.2}$   & 0.20$^{+0.00}_{-0.14}$  & 0.6 &   46.3 &   53.9 &   54.4 &   54.5 &   34.6 &   - &   34.8 \\
 NGC4656-2 & 36.4$^{+-3.1}_{-30.5}$ & 0.45$^{+0.00}_{-0.35}$  & 0.6 &   48.0 &   53.5 &   54.7 &   54.7 &   35.0 &     &   39.5 \\
 NGC4449-1 & 46.1$^{+4.0}_{-0.0}$   & 0.31$^{+0.07}_{-0.05}$  & 0.6 &   47.5 &   53.3 &   54.6 &   54.6 &   34.4 &   - &   34.4 \\
 NGC5253-2 & 49.1$^{+0.0}_{-38.5}$  & 0.25$^{+0.00}_{-0.19}$  & 0.6 &   47.3 &   53.2 &   54.5 &   54.5 &   34.2 &   - &   34.2 \\
 NGC5253-1 & 49.1$^{+0.1}_{-0.1}$   & 0.50$^{+0.05}_{-0.08}$  & 0.7 &   47.6 &   53.5 &   54.8 &   54.8 &   34.5 &   - &   34.5 \\
\hline
\end{tabular}
\caption{Stellar feedback of the CLUES targets measured for the young stellar population (top table) and for the old stellar population (bottom panel). Column (1): source name. Columns (2): age of the stellar population from Paper I. Columns (3): mass of the stellar population from Paper I. Column (4): light flux ratio at 1270 Å of the single stellar population relative to the total population. Column (5): photon production rate. Column (6), (7) and (8): mechanical energy produced by stellar winds, SNe and their sum. Column (9), (10) and (11): mechanical luminosity produced by stellar winds, SNe and their sum. We have not propagated the errors on age, metallicity, and mass into errors on the photoionizing flux, energy and luminosity.} These values are based on the models presented in Paper I and carry large uncertainty (of roughly 1 dex) from the interpolation of Starburst99 libraries and best-fit parameters of the spectral fit.
\label{tab:feedback_YO}
\end{table*}

The total photon production rate $Q_H$ ranges between the order of $10^{50}$ and $10^{52}$ photons/s for most targets, which provides a radiative flux that is likely able to not only ionize the gas clouds in the ISM surrounding the cluster, but also power gas outflows (see Section \ref{sec:out_correlations}). When comparing the two stellar populations of each target, it is clear that the young population dominates the production rate of ionizing photons by several orders of magnitude. For the total mechanical energy, we find a mean logarithmic value of log$(E_{\rm Tot}/\rm erg) = 54.4 \pm 0.9$ (where the error is the standard deviation), with the old stellar component contributing about two orders of magnitude more than the young one, since this is a quantity integrated over time. For the total mechanical luminosity log$(L_{\rm Tot}/\rm erg\,s^{-1}) = 39.3 \pm 0.8$, with the young stellar population dominating due to the power in their stellar winds that end after about 6 Myr. This amount of energy powers gas outflows of a few hundred km~s$^{-1}$, which we detect in the majority of our targets (see Section \ref{sec:outflows}).
For the SNe, the mechanical luminosity is zero at ages below 4 Myr (as no SN has exploded yet), it dominates over the stellar wind feedback after the age of about 10 Myr, and it drops to zero at around 40 Myr as SNe stop exploding. We note that the Starburst99 models do not include the effect of binary stars, which might prolong the period in which SNe explode \citep{Eldridge2009,Gotberg2017}.

\section{Kinematics of the interstellar gas} \label{sec:kinematics}

FUV photons can be absorbed by different ions, with different ionization potentials.
Therefore, COS FUV spectroscopy offers a powerful tool to detect multiple phases of interstellar gas and to measure their kinematics as well as the column density. In this section, we report the employed methodology and the derived physical properties of the intervening ISM.

\subsection{Spectral resampling and normalization}
\begin{figure}
    \centering
    \includegraphics[width=\columnwidth]{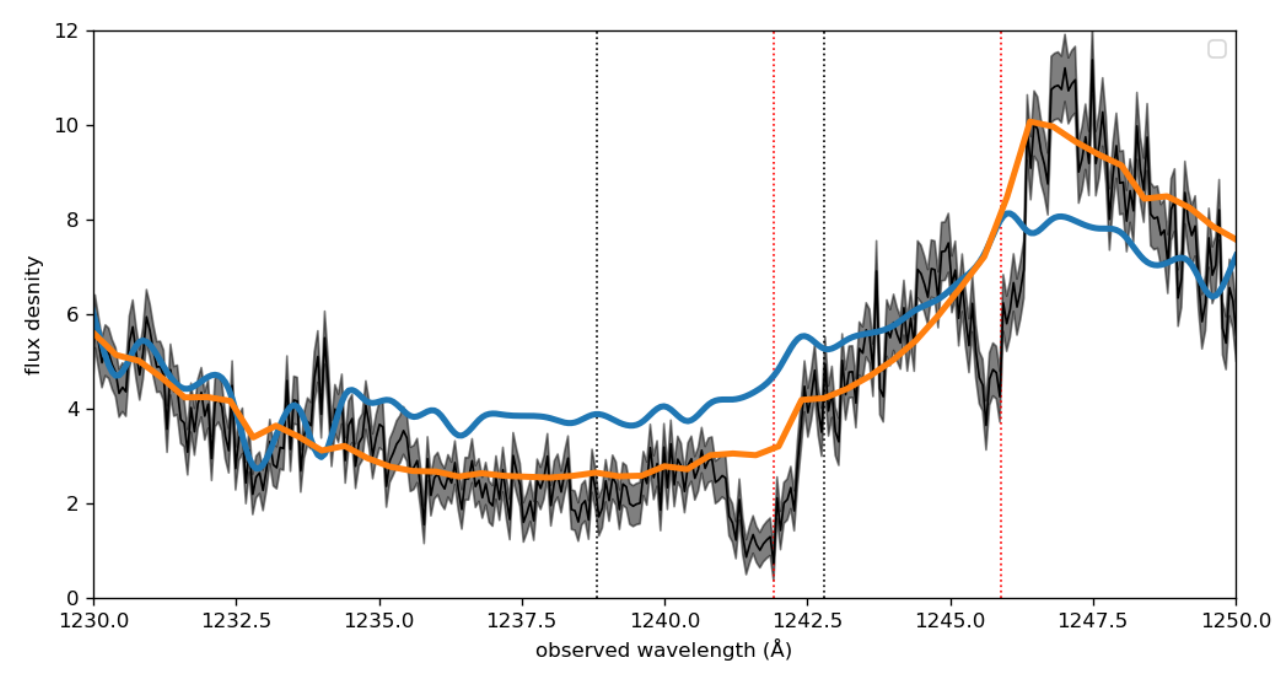}
     \caption{\NV$\,$ P-Cygni line of the target M95-YSC1 (black line). In blue the best-fit model as reported in Paper I. In orange the improved model found by fitting only between 1230 and 1250 Å, rather than the whole available spectrum. The red dashed vertical lines mark the wavelengths where the \NV$\,$ ISM absorption is expected due to the gas in the target galaxy, whereas the black dashed lines mark the Milky Way gas absorption.}
    \label{fig:new_mod}
\end{figure}

We first resampled the spectra to the nominal resolution element 0.06 \AA, which is the resolution intrinsic to the COS spectrograph. The spectra used in this step are the ones corrected for Milky Way reddening and for \HI$\,$ absorption (Paper I). We then resampled the best-fit models, originally binned at 0.4 Å, to the same binning of the data 0.06 \AA. To do so, we used a cubic spline interpolation to approximate the model and extrapolate the values of the returned function for the denser grid of wavelengths. Next, we normalized the data so that the continuum level is equal to 1, i.e. we divided the data by the best-fit models. For a few targets, the best-fit model has large residuals (up to 30 \% \footnote{residuals measured at the peak of the P-Cygni lines}) at the wavelengths around the absorption lines listed in Table \ref{tab:ISM_lines}. This is likely due to limitations of the models that do not account for binary and multiple systems of stars, rotating stars and the formation of very massive stars exceeding the upper limit of the IMF (see Paper I). Possibly, issues in the stellar atmosphere, wind models as well as the differential extinction of the young and old stellar population might also drive the residuals. For this reason, we repeated for these targets the stellar population fits with the same method as described in Paper I and the same parameters but limiting the fit to the spectral window of individual lines, rather than the full extent of the spectrum. Figure \ref{fig:new_mod} shows an example of how the model for the \NV$\,$line of the target M95-YSC1 is improved. The list of targets with the individually remodeled lines are: M74-YSC2 for \SiIV, \SiII, \CIV; M95-YSC1 for \NV, \SiIV, \SiII; NGC1512-YSC2 for \SiIV, \SiII, \CIV, \SII; for NGC1566-YSC2 \SiIV, \SiII, \CII; NGC5253-YSC2 for \SiIV, \SiII, \CIV, \AlII. We did not update the physical parameters of the stellar populations with the values of these new fits, as they cover only a very narrow portion of the spectrum (about 20 Å). 

\begin{table}
    \centering
    \begin{tabular}{lrrr}
    \hline
    ion species & $\lambda_{vac}$ (\AA) & $f_{lu}$ & IP (eV) \\
    (1) & (2) & (3) & (4) \\    
\hline
 \AlII     & 1670.79 & 1.70 & 6.0 \\
 Fe II     & 1608.45 & 0.06 & 7.9 \\
 \SiII     & 1190.42 & 0.28 & 8.2 \\
 \SiII     & 1193.29 & 0.59 & 8.2 \\
 \SiII     & 1260.42 & 1.45 & 8.2 \\
 \SiII     & 1304.47 & 0.11 & 8.2 \\
 \SiII     & 1526.72 & 0.13 & 8.2 \\
 \SII      & 1250.58 & 0.01 & 10.4 \\
 \SII      & 1253.81 & 0.01 & 10.4 \\
 \SII      & 1259.52 & 0.02 & 10.4 \\
 \CII     & 1334.53 & 0.13 & 11.3 \\
 \CII      & 1335.71 & 0.12 & 11.3 \\
 \SiIV     & 1393.75 & 0.54 & 33.5 \\
 \SiIV     & 1402.77 & 0.03 & 33.5 \\
 \CIV      & 1548.20 & 0.19 & 47.9 \\
 \CIV      & 1550.78 & 0.10 & 47.9 \\
 \NV       & 1238.82 & 0.16 & 77.5 \\
 \NV       & 1242.80 & 0.08 & 77.5 \\
\hline
    \end{tabular}
    \caption{List of FUV absorption lines that we detected in the CLUES data, which are produced by metals in the ISM. Column (1): name of the ion species; Column (2): vacuum wavelength of the electronic transition; Column (3): oscillator strength, proxy of the absorption probability; Column (4): ionization potential of the ion species. Atomic data from \cite{Jitrik2004}.}
    \label{tab:ISM_lines}
\end{table}

\subsection{Zero-velocity of the gas}
\label{sec:redshifts}

In this study we are mainly interested in the kinematics of the gas and more specifically in how the gas flows as an effect of the star cluster feedback (radiation, momentum and energy released by massive stars, stellar winds and SNe). We, therefore, want to measure the gas velocity relative to the systemic velocity of the targeted star cluster. This requires that we know the velocity of the stellar component to a precision better than done in Paper I for the stellar population fitting. 

In this work we measured the zero-velocity using the \CIII$\,$1247 photospheric absorption line in the FUV, arising in the photosphere of OB stars in the target clusters. When this line was not available, we used instead the other photospheric line \CIII$\,$1176\footnote{Another photospheric line that is in principle available in the spectral window of our data is \SV$\,$1501.76, however we marginally detected this line only in a few targets.}, although it is contaminated by the stellar wind component and could present a P-Cygni profile. Both \CIII$\,$ absorption features are in reality the result of multiple transitions of different strengths that blend with each other. For this reason we compared the redshift values found with other measurements that we obtained using alternative methods. 

We used the H$\alpha$ emission line extracted at the position of the cluster (i.e., \HII$\,$ region powered by the cluster) from MUSE data, which are available for M95-YSC1, NGC1566-YSC1, NGC1566-YSC2, M74-YSC2, NGC1512-YSC2, NGC1313-YSC2. For these targets, we compared the H$\alpha$ based redshift with the \CIII$\,$ based values and found that the \CIII $\,1247.38$ measurements are in agreement (absolute velocity difference less than 30 km/s) while \CIII $\,1175.99$ measurements have an absolute velocity difference above 60 km/s. Following the results of this comparison, we then used \CIII $\,1247.38$ also for M51-YSC1, for M51-YSC2 and M74-YSC1, that have no MUSE data. 
For the four remaining targets, with no detection of \CIII $\,1247.38$ and no MUSE data, we used the galactic systemic velocity, taken from the NED database. These values were derived from the following \HI$\,$ surveys: \cite{Wolfinger2013} for NGC4485-YSC1 and NGC4485-YSC2, \cite{Devaucouleurs} for NGC4656-YSC1 and NGC4656-YSC2, NGC5253-YSC1, NGC5253-YSC2 and \cite{Koribalski2004} for NGC1313-YSC1. We note that these galaxies are expected to suffer less from galaxy rotation since they are dwarf galaxies, or present some irregularity in their morphology. 

We report in Table \ref{tab:redshifts} the values of redshifts obtained with the different datasets for each target, marking in boldface the value that we adopted. Finally, we performed a cross correlation between best fit stellar population model and data (masking the ISM features) to sanity check the redshifts based on \HI$\,$ and we found good agreement (absolute velocity difference less than 23 km/s) except for the target NGC4656-YSC1. For this cluster we found a smaller redshift, with a velocity difference of 73 km/s, which would make the detected inflow a faster inflow. This final test confirms that our results are robust against our selected zero-velocities.

\begin{table}
\centering
\begin{tabular}{llllr}
\hline
target & \CIII $\,1247$ & \CIII $\,1176$ & H$\alpha$ & \HI\\
\hline
 M51-1     & \textbf{0.00148}  & -        & -        & 0.00131  \\
 M51-2     & \textbf{0.001319} & 0.000808 & -        & 0.00131  \\
 M74-1     & \textbf{0.002252} & -        & -        & 0.002188 \\
 M74-2     & \textbf{0.002228} & -        & 0.002157 & 0.002188 \\
 M95-1     & \textbf{0.002401} & -        & 0.002488 & 0.002595 \\
 NGC1313-1 & -        & 0.001379 & -        & \textbf{0.00157}  \\
 NGC1313-2 & -        & 0.001346 & \textbf{0.001572} & 0.00157  \\
 NGC1512-1 & \textbf{0.003137} & -        & -        & 0.00299  \\
 NGC1512-2 & \textbf{0.002710} & -        & 0.003036 & 0.00299  \\
 NGC1566-1 & \textbf{0.005160} & 0.004674 & 0.005119 & 0.00502  \\
 NGC1566-2 & \textbf{0.005055} & 0.003806 & 0.005041 & 0.00502  \\
 NGC4485-1 & -        & 0.001201 & -        & \textbf{0.00164} \\
 NGC4485-2 & -        & 0.001256 & -        & \textbf{0.00164} \\
 NGC4656-1 & -        & 0.001864 & -        & \textbf{0.002142*} \\
 NGC4656-2 & \textbf{0.002151} & 0.001674 & -        & 0.002142 \\
 NGC5253-1 & \textbf{0.001407} & -        & -        & 0.00136  \\
 NGC5253-2 & \textbf{0.001446} & -        & -        & 0.00136  \\
 NGC7793-1 & \textbf{0.000768} & -        & 0.000891 & 0.00076  \\
 NGC7793-2 & \textbf{0.001007} & -        & 0.000969 & 0.00076  \\
\hline
\end{tabular}
\caption{Redshift measurements based on different lines: photospheric absorption lines \CIII $\,1247$ and \CIII $\,1176$ when detected, H$\alpha$ line from MUSE data when available and galactic measurements from the literature, typically part of \HI$\,$ surveys (see references in the text). Marked in boldface are the values that we adopted for our analysis of the gas kinematics. The asterisk serves as a reminder that the redshift value of NGC4656-YSC1 might be smaller, according to the cross correlation between data and model.}
\label{tab:redshifts}
\end{table}

\subsection{Multi-component Voigt fit}
\label{sec:voigt}

For the analysis of the ISM gas kinematics, we modeled the absorption lines listed in Table \ref{tab:ISM_lines} with a Voigt profile using the python tool VoigtFit \citep[][]{Krogager2018}. This software allowed us to fit any of the lines that are detected in the data simultaneously, as well as to tie the Voigt parameters separately for low ($<13.6$ eV) and high ($>13.6$ eV) ionization lines (see the ionization potential values in Table \ref{tab:ISM_lines}). We will refer to the low-ionization and high-ionization lines as ``neutral and ionized gas" hereafter. A Voigt function has three parameters: the column density of the specific ion species $N_{ion}$, the central velocity (where the minimum of the function is) relative to the zero velocity, the Doppler parameter, $b$, that measures the width of the absorption profile and is a function of the gas temperature and turbulence. We assume here a covering fraction of 1 and we justify this assumption in Sec \ref{sec:covering}. We tied the velocity structure (central velocity and $b$ parameter) across all species of ions, separately for the low and high ionization lines, while we let free the column density of each component of each species. In tying the $b$ parameter across different ions, we assumed that the broadening of the absorption lines is dominated by turbulence (see Section \ref{sec:voigt_results}) rather than by thermal motion, in which case it would scale inversely  with the ion mass. We chose the zero velocity (systemic redshift) as described in Section \ref{sec:redshifts}. We also tied the column densities of all the different transitions of the same ion. The Voigt model is convolved with the instrumental LSF, as well as the broadening introduced by the extent of the source, which we measured in the NUV acquisition image (see Table \ref{tab:res}). 

We performed a multiple-component Voigt fit in order to model the absorption by the interstellar gas, finding components of gas outflowing with velocities of up to a few hundred km~s$^{-1}$. We fixed the first component to the zero velocity in all targets to account for stationary gas, whereas we let the velocity of the other components free. This method is based on the approach by \cite{Xu2022} with two main differences: (i) our model profile for each component is a Voigt function rather than a Gaussian; (ii) we allowed for the number of extra components (in addition to the stationary gas) to be larger than one. The optimal number of components necessary to model the observed line profile is found through a systematic analysis based on the Bayesian information criterion (BIC)\footnote{For a formal definition of the estimator and its applications in astrophysical models see \cite{Liddle2007}.}. We used a series of models with an increasing number of components and calculated the BIC estimator for each model. By selecting the model with the lowest BIC values, we found the model with the minimum number of components that best describes the data. We typically find two components for each transition, one tracing stationary gas and one tracing outflowing gas.

\subsubsection{Correction for Milky Way and Magellanic System absorption}
Because the gas in the Milky Way lies along the line of sight to the CLUES targets, we detected additional absorption components associated with it. While for the high ionization lines, the Milky Way components are distinguishable from the target components and we could ignore them, this was not the case for some low ionization lines close to each other (e.g. \SII$\,$ 1250,1253,1259 and \SiII$\,$ 1260). We modeled accurately the Milky Way components of low ionization species by including in the fit lines such as \AlII$\,$ 1670, \SiII$\,$ 1526, which are not blended with each other. By having a fixed velocity structure, we could de-blend the Milky Way components for all lines in the low ionization regime. 
We found an additional source of contamination consisting in absorption components due to the gas in the Magellanic System, which lies along the line of sight to the targets NGC7793-YSC1, NGC7793-YSC2, NGC1313-YSC1, NGC1313-YSC2, NGC5253-YSC1, NGC5253-YSC2 as mentioned in Section \ref{sec:data}. As shown in Appendix, we have been able to disentangle the components associated with the Magellanic System and could study the gas in the target galaxies without contamination.

\subsubsection{Line saturation and resonance scattering}
We examined the potential issue of line saturation that can affect the absorption profiles \cite[e.g.][]{James2014}. Classical line saturation is the scenario in which the depth of the absorption line reaches the zero flux level and therefore ceases to probe any additional absorbing gas. We checked for the presence of line saturation in our data for \SiII, which we used for measuring the column density of the neutral gas. We visually observed saturation in four targets, i.e. NGC4485-YSC1, NGC4485-YSC2, NGC4656-YSC1, NGC4656-YSC2 (see Appendix), however only at a marginal level. For these targets, our measurement of the column densities may be slightly underestimated. Another type of saturation occurs with partial coverage, when multiple lines of sight trace both large and small column densities of gas. The result is that the absorption line appears unsaturated, when in reality it is. It is possible to unveil hidden saturation in data by looking at how well a model with a given column density reproduces the depth of multiple lines of the same ion of varying oscillator strength. In data suffering from hidden saturation, the lines with the highest oscillator strength values appear weaker and are overpredicted by the model. We checked this for the triplet \SII$\,$ 1250,1253,1259 Å, detected in all targets, and we did not find traces of hidden saturation. We notice that hidden saturation is a function of the column density of an ion and might still exist for \SiII$\,$. However, we can conclude that the presence of hidden saturation is not important, since we found no significant partial coverage with the analysis on the covering fraction (see Section \ref{sec:covering}).

We looked for any signature of resonance scattering that could contaminate the absorption profile of the FUV transitions of interest. This phenomenon is also known as infilling effect and described in, e.g., \cite{Heckman2015,Xu2022}. As a signature of resonance scattering, we searched for the associated fluorescent emission lines \SiII* 1265 and \SiII* 1532. Since we do not detect any fluorescent line, we conclude that the infilling effect is negligible in our data. We note that the lack of resonance scattering suggests that the neutral gas wind radius is larger than the COS aperture \citep[see figures 6 and 7 in ][]{Zhu2015}, which is consistent with the outflow radii estimated in Section \ref{sec:outflows}.

\subsubsection{Results of the Voigt fits}
\label{sec:voigt_results}
Table \ref{tab:voigtfit} and \ref{tab:voigtfit2} list for each target all the ions that were detected and fitted as well as the best-fit values of central velocity, $b$ parameter and column density for each absorption component detected. Errors returned by the software VoigtFit are listed for all parameters. The derived column density values are associated to unphysically large errors in the following cases: NGC1566-YSC1 (stationary component), NGC1313-YSC2 \SII$\,$ (outflow component), NGC4656-YSC2 \CIV, \SiIV$\,$ (stationary component), NGC5253-YSC1 \CII$\,$ (outflow component). For these cases, we reported the column densities without uncertainties. These values should be considered as upper limits, since they refer to absorption components that are too shallow to be constrained.
We detected at least one outflow component in all cases, except NGC4485-YSC2. In two cases, we detected two outflowing components, in the neutral gas phase for NGC1566-YSC1 and in the ionized gas phase for NGC4656-YSC2. In five targets (NGC1313-YSC2, NGC4656-YSC1, M51-YSC2, NGC4656-YSC2, NGC5253-YSC2) we detected an inflow of gas (positive velocities) rather than an outflow, which we discuss in Section \ref{sec:discussion}.
\begin{table*}
\centering
\begin{tabular}{lcccccccccc}
\hline
target & ions & $b_0$ & $log(N_0\rm [cm\,s^{-1}])$ & $v_1$ &  $b_1$ & $log(N_1\rm [cm\, s^{-1}])$ & $v_2$ & $b_2$ & $log(N_2\rm [cm\,s^{-1}])$\\

 & & [km~s$^{-1}$] &  & [km~s$^{-1}$] & [km~s$^{-1}$] & & [km~s$^{-1}$] & [km~s$^{-1}$] & \\
 (1) & (2) & (3) & (4) & (5) & (6) & (7) & (8) & (9) & (10))\\
\hline
  M74-YSC2     & AlII & 46$ \pm $6   & 13.7$ \pm $0.4   & -42$ \pm $14  & 27$ \pm $7   & 15.1$ \pm $1.3    &               &            &                \\
              & CII  & 46$ \pm $0   & 15.2$ \pm $0.4   & -42$ \pm $0   & 27$ \pm $0   & 18.1$ \pm $0.7    &               &            &                \\
              & CIV  & 54$ \pm $22  & 14.1$ \pm $1.5   & -46$ \pm $71  & 72$ \pm $47  & 14.8$ \pm $0.6    &               &            &                \\
              & SII  & 46$ \pm $6   & 15.3$ \pm $0.1   & -42$ \pm $14  & 27$ \pm $7   & 14.8$ \pm $0.5    &               &            &                \\
              & SiII & 46$ \pm $6   & 14.8$ \pm $0.2   & -42$ \pm $14  & 27$ \pm $7   & 16.2$ \pm $0.7    &               &            &                \\
              & SiIV & 54$ \pm $22  & 12.8$ \pm $3.9   & -46$ \pm $71  & 72$ \pm $47  & 14.1$ \pm $0.6    &               &            &                \\
 M95-YSC1     & AlII & 65$ \pm $4   & 13.5$ \pm $0.1   & -115$ \pm $12 & 82$ \pm $8   & 13.6$ \pm $0.1    &               &            &                \\
              & CII  & 65$ \pm $0   & 14.5$ \pm $0.6   & -115$ \pm $0  & 82$ \pm $0   & 15.6$ \pm $0.2    &               &            &                \\
              & NV   & 74$ \pm $22  & 14.3$ \pm $0.2   & -74$ \pm $45  & 111$ \pm $28 & 14.4$ \pm $0.3    &               &            &                \\
              & SII  & 65$ \pm $4   & 15.4$ \pm $0.1   & -115$ \pm $12 & 82$ \pm $8   & 15.0$ \pm $0.1    &               &            &                \\
              & SiII & 65$ \pm $4   & 14.8$ \pm $0.1   & -115$ \pm $12 & 82$ \pm $8   & 14.9$ \pm $0.1    &               &            &                \\
              & SiIV & 74$ \pm $22  & 13.7$ \pm $1.2   & -74$ \pm $45  & 111$ \pm $28 & 14.5$ \pm $0.2    &               &            &                \\
 M74-YSC1     & AlII & 34$ \pm $22  & 13.6$ \pm $0.6   & -72$ \pm $5   & 74$ \pm $4   & 13.5$ \pm $0.1    &               &            &                \\
              & CIV  & 185$ \pm $47 & 14.4$ \pm $0.1   & -93$ \pm $9   & 74$ \pm $11  & 14.8$ \pm $0.1    &               &            &                \\
              & SII  & 34$ \pm $22  & 14.9$ \pm $0.3   & -72$ \pm $5   & 74$ \pm $4   & 15.2$ \pm $0.1    &               &            &                \\
              & SiII & 34$ \pm $22  & 14.4$ \pm $0.6   & -72$ \pm $5   & 74$ \pm $4   & 15.0$ \pm $0.1    &               &            &                \\
              & SiIV & 185$ \pm $47 & 14.0$ \pm $0.1   & -93$ \pm $9   & 74$ \pm $11  & 14.3$ \pm $0.1    &               &            &                \\
 NGC1512-YSC2 & CII  & 25$ \pm $0   & 13.9$ \pm $0.3   & -62$ \pm $0   & 26$ \pm $0   & 15.7$ \pm $0.5    &               &            &                \\
              & SII  & 25$ \pm $6   & 14.8$ \pm $0.1   & -62$ \pm $3   & 26$ \pm $4   & 14.7$ \pm $0.1    &               &            &                \\
              & SiII & 25$ \pm $6   & 13.3$ \pm $0.1   & -62$ \pm $3   & 26$ \pm $4   & 14.7$ \pm $0.2    &               &            &                \\
              & SiIV & 68$ \pm $658 & 11.5$ \pm $7.2   & *-186$ \pm $30 & *167$ \pm $47 & *14.0$ \pm $0.1    &               &            &                \\
 M51-YSC1     & AlII & 64$ \pm $4   & 13.6$ \pm $0.1   &   &    &     &               &            &                \\
              & CIV  & 70$ \pm $13  & 15.4$ \pm $0.2   & -193$ \pm $17 & 46$ \pm $28  & 14.2$ \pm $0.2    &               &            &                \\
              & SII  & 64$ \pm $4   & 15.1$ \pm $0.1   &  &   &    &               &            &                \\
              & SiII & 64$ \pm $4   & 15.1$ \pm $0.1   &   &   &    &               &            &                \\
 NGC1566-YSC2 & SiII & 338$ \pm $54 & 14.5$ \pm $0.1   & -141$ \pm $13 & 40$ \pm $23  & 13.7$ \pm $0.1    &               &            &                \\
              & SiIV & 577$ \pm $99 & 14.3$ \pm $0.0   & -130$ \pm $14 & 187$ \pm $17 & 14.1$ \pm $0.1    &               &            &                \\
 NGC7793-YSC1 & AlII & 35$ \pm $4   & 13.3$ \pm $0.2   & -45$ \pm $10  & 43$ \pm $6   & 13.4$ \pm $0.2    &               &            &                \\
              & CII  & 35$ \pm $0   & 15.2$ \pm $4.0   & -45$ \pm $0   & 43$ \pm $0   & 15.0$ \pm $0.6    &               &            &                \\
              & CIV  & 26$ \pm $8   & 14.6$ \pm $0.1   & -43$ \pm $15  & 77$ \pm $9   & 14.4$ \pm $0.1    &               &            &                \\
              & FeII & 35$ \pm $4   & 14.6$ \pm $0.2   & -45$ \pm $10  & 43$ \pm $6   & 14.6$ \pm $0.2    &               &            &                \\
              & SII  & 35$ \pm $4   & 15.2$ \pm $0.1   & -45$ \pm $10  & 43$ \pm $6   & 15.3$ \pm $0.1    &               &            &                \\
              & SiII & 35$ \pm $4   & 14.3$ \pm $0.2   & -45$ \pm $10  & 43$ \pm $6   & 14.8$ \pm $0.2    &               &            &                \\
              & SiIV & 26$ \pm $8   & 13.6$ \pm $0.2   & -43$ \pm $15  & 77$ \pm $9   & 14.0$ \pm $0.1    &               &            &                \\
 NGC4485-YSC2 & AlII & 62$ \pm $2   & 14.0$ \pm $0.1   &               &              &                   &               &            &                \\
              & FeII & 62$ \pm $2   & 15.3$ \pm $0.1   &               &              &                   &               &            &                \\
              & SII  & 62$ \pm $2   & 15.8$ \pm $0.0   &               &              &                   &               &            &                \\
              & SiII & 62$ \pm $2   & 15.3$ \pm $0.1   &               &              &                   &               &            &                \\
              & SiIV & 118$ \pm $12 & 14.0$ \pm $0.0   &               &              &                   &               &            &                \\
 NGC4485-YSC1 & AlII & 57$ \pm $4   & 14.4$ \pm $0.3   &               &              &                   &               &            &                \\
              & CII  & 57$ \pm $0   & 16.9$ \pm $0.5   &               &              &                   &               &            &                \\
              & CIV  & 89$ \pm $6   & 14.5$ \pm $0.0   & -82$ \pm $4   & 35$ \pm $5   & 14.5$ \pm $0.1    &               &            &                \\
              & FeII & 57$ \pm $4   & 15.2$ \pm $0.1   &               &              &                   &               &            &                \\
              & SII  & 57$ \pm $4   & 15.4$ \pm $0.0   &               &              &                   &               &            &                \\
              & SiII & 57$ \pm $4   & 15.7$ \pm $0.3   &               &              &                   &               &            &                \\
              & SiIV & 89$ \pm $6   & 14.1$ \pm $0.0   & -82$ \pm $4   & 35$ \pm $5   & 14.2$ \pm $0.1    &               &            &                \\
\hline
\end{tabular}
\caption{See caption of Table \ref{tab:voigtfit2} (continuation).}
\label{tab:voigtfit}
\end{table*}
\begin{table*}
\centering
\begin{tabular}{lcccccccccc}
\hline
target & ions & $b_0$ & $log(N_0\rm [cm\,s^{-1}])$ & $v_1$ &  $b_1$ & $log(N_1\rm [cm\, s^{-1}])$ & $v_2$ & $b_2$ & $log(N_2\rm [cm\,s^{-1}])$\\

 & & [km~s$^{-1}$] &  & [km~s$^{-1}$] & [km~s$^{-1}$] & & [km~s$^{-1}$] & [km~s$^{-1}$] & \\
 (1) & (2) & (3) & (4) & (5) & (6) & (7) & (8) & (9) & (10))\\
\hline
 NGC1313-YSC1 & AlII & 28$ \pm $3   & 14.9$ \pm $0.5   & -48$ \pm $0   & 71$ \pm $8   & 13.2$ \pm $0.1    &               &            &                \\
              & FeII & 28$ \pm $3   & 15.0$ \pm $0.2   & -48$ \pm $0   & 71$ \pm $8   & 14.3$ \pm $0.2    &               &            &                \\
              & SII  & 28$ \pm $3   & 15.4$ \pm $0.1   & -48$ \pm $0   & 71$ \pm $8   & 15.0$ \pm $0.1    &               &            &                \\
              & SiII & 28$ \pm $3   & 16.3$ \pm $0.4   & -48$ \pm $0   & 71$ \pm $8   & 14.6$ \pm $0.1    &               &            &                \\
              & SiIV & 20$ \pm $22  & 13.9$ \pm $0.3   & -30$ \pm $18  & 94$ \pm $23  & 14.2$ \pm $0.1    &               &            &                \\
 NGC1566-YSC1 & AlII & 7$ \pm $73   & 8.8 & -64$ \pm $11  & 55$ \pm $7   & 13.7$ \pm $0.2    & *-222$ \pm $7  & *15$ \pm $7 & *12.6$ \pm $0.6 \\
              & SiII & 7$ \pm $73   & 15.1 & -64$ \pm $11  & 55$ \pm $7   & 14.8$ \pm $0.2    & *-222$ \pm $7  & *15$ \pm $7 & *14.1$ \pm $0.5 \\
              & SiIV & 5$ \pm $593  & 13.5 & -122$ \pm $25 & 171$ \pm $26 & 14.5$ \pm $0.1    &               &            &                \\
 NGC7793-YSC2 & AlII & 17$ \pm $5   & 13.5$ \pm $0.8   & -38$ \pm $19  & 24$ \pm $8   & 13.3$ \pm $0.5    &               &            &                \\
              & CIV  & 23$ \pm $12  & 13.3$ \pm $0.6   & -84$ \pm $10  & 59$ \pm $17  & 14.2$ \pm $0.1    &               &            &                \\
              & FeII & 17$ \pm $5   & 14.6$ \pm $0.5   & -38$ \pm $19  & 24$ \pm $8   & 14.5$ \pm $0.3    &               &            &                \\
              & SII  & 17$ \pm $5   & 15.1$ \pm $0.2   & -38$ \pm $19  & 24$ \pm $8   & 14.9$ \pm $0.3    &               &            &                \\
              & SiII & 17$ \pm $5   & 14.6$ \pm $0.6   & -38$ \pm $19  & 24$ \pm $8   & 15.1$ \pm $0.8    &               &            &                \\
              & SiIV & 23$ \pm $12  & 13.7$ \pm $0.2   & -84$ \pm $10  & 59$ \pm $17  & 13.9$ \pm $0.1    &               &            &                \\
 NGC1313-YSC2 & CIV  & 145$ \pm $77 & 14.3$ \pm $0.1   & -28$ \pm $6   & 31$ \pm $12  & 14.2$ \pm $0.2    &               &            &                \\
              & FeII & 42$ \pm $3   & 14.8$ \pm $0.1   & 56$ \pm $11   & 10$ \pm $18  & 15.5$ \pm $5.3    &               &            &                \\
              & SII  & 42$ \pm $3   & 15.6$ \pm $0.1   & 56$ \pm $11   & 10$ \pm $18  & 12.0  &               &            &                \\
              & SiII & 42$ \pm $3   & 15.0$ \pm $0.2   & 56$ \pm $11   & 10$ \pm $18  & 14.5$ \pm $3.0    &               &            &                \\
              & SiIV & 145$ \pm $77 & 13.9$ \pm $0.1   & -28$ \pm $6   & 31$ \pm $12  & 14.1$ \pm $0.1    &               &            &                \\
 NGC4656-YSC1 & AlII & 39$ \pm $3   & 13.7$ \pm $0.2   & 73$ \pm $4    & 34$ \pm $4   & 13.3$ \pm $0.2    &               &            &                \\
              & CII  & 39$ \pm $0   & 16.0$ \pm $0.3   & 73$ \pm $0    & 34$ \pm $0   & 15.8$ \pm $0.3    &               &            &                \\
              & FeII & 39$ \pm $3   & 14.7$ \pm $0.1   & 73$ \pm $4    & 34$ \pm $4   & 14.8$ \pm $0.2    &               &            &                \\
              & SII  & 39$ \pm $3   & 15.7$ \pm $0.1   & 73$ \pm $4    & 34$ \pm $4   & 15.8$ \pm $0.1    &               &            &                \\
              & SiII & 39$ \pm $3   & 14.8$ \pm $0.1   & 73$ \pm $4    & 34$ \pm $4   & 14.5$ \pm $0.1    &               &            &                \\
              & SiIV & 140$ \pm $13 & 14.1$ \pm $0.0   &               &              &                   &               &            &                \\
 M51-YSC2     & AlII & 69$ \pm $4   & 13.6$ \pm $0.1   & *-162$ \pm $4  & *31$ \pm $5   & *13.1$ \pm $0.1    &               &            &                \\
              & CIV  & 150$ \pm $8  & 14.7$ \pm $0.0   & *-147$ \pm $5  & *48$ \pm $8   & *14.8$ \pm $0.1    &               &            &                \\
              & SiII & 69$ \pm $4   & 15.0$ \pm $0.1   & *-162$ \pm $4  & *31$ \pm $5   & *14.0$ \pm $0.1    &               &            &                \\
              & SiIV & 150$ \pm $8  & 14.4$ \pm $0.0   & *-147$ \pm $5  & *48$ \pm $8   & *13.8$ \pm $0.1    &               &            &                \\
 NGC4656-YSC2 & AlII & 64$ \pm $3   & 13.7$ \pm $0.1   & 75$ \pm $5    & 33$ \pm $7   & 13.7$ \pm $0.2    &               &            &                \\
              & CIV  & 8$ \pm $60   & 9.4  & -41$ \pm $3   & 15$ \pm $8   & 14.5$ \pm $0.5    & -108$ \pm $10 & 86$ \pm $9 & 14.3$ \pm $0.1 \\
              & FeII & 64$ \pm $3   & 14.9$ \pm $0.1   & 75$ \pm $5    & 33$ \pm $7   & 14.8$ \pm $0.2    &               &            &                \\
              & SII  & 64$ \pm $3   & 15.5$ \pm $0.1   & 75$ \pm $5    & 33$ \pm $7   & 15.6$ \pm $0.1    &               &            &                \\
              & SiII & 64$ \pm $3   & 15.3$ \pm $0.1   & 75$ \pm $5    & 33$ \pm $7   & 14.1$ \pm $0.5    &               &            &                \\
              & SiIV & 8$ \pm $60   & 10.0  & -41$ \pm $3   & 15$ \pm $8   & 14.2$ \pm $0.6    & -108$ \pm $10 & 86$ \pm $9 & 14.1$ \pm $0.1 \\
 NGC5253-YSC2 & AlII & 165$ \pm $64 & 12.8$ \pm $0.3   & 33$ \pm $4    & 60$ \pm $6   & 13.2$ \pm $0.1    &               &            &                \\
              & SII  & 165$ \pm $64 & 14.7$ \pm $0.3   & 33$ \pm $4    & 60$ \pm $6   & 15.4$ \pm $0.1    &               &            &                \\
              & SiII & 165$ \pm $64 & 14.3$ \pm $0.1   & 33$ \pm $4    & 60$ \pm $6   & 14.5$ \pm $0.1    &               &            &                \\
              & SiIV & 49$ \pm $5   & 14.1$ \pm $0.0   & -36$ \pm $17  & 344$ \pm $56 & 14.3$ \pm $0.0    &               &            &                \\
 NGC5253-YSC1 & AlII & 45$ \pm $2   & 13.5$ \pm $0.1   & -84$ \pm $3   & 13$ \pm $3   & 12.7$ \pm $0.3    &               &            &                \\
              & CII  & 45$ \pm $0   & 15.6$ \pm $0.2   & -84$ \pm $0   & 13$ \pm $0   & 12.3  &               &            &                \\
              & SII  & 45$ \pm $2   & 15.5$ \pm $0.0   & -84$ \pm $3   & 13$ \pm $3   & 14.4$ \pm $0.3    &               &            &                \\
              & SiII & 45$ \pm $2   & 14.7$ \pm $0.1   & -84$ \pm $3   & 13$ \pm $3   & 15.6$ \pm $0.8    &               &            &                \\
              & SiIV & 6$ \pm $8    & 16.0$ \pm $3.9   & -47$ \pm $8   & 148$ \pm $8  & 14.3$ \pm $0.0    &               &            &                \\
\hline
\end{tabular}
\caption{Results of the Voigt fits to the FUV absorption lines in the CLUES sample. (1) Name of the target cluster. (2) Name of the ions whose transitions are fitted with a Voigt profile. (3) $b$ parameter of the zero-velocity gas component. (4) Logarithm of the column density of the zero-velocity gas component. (5) Central velocity of the first outflow component. (6) $b$ parameter of the first outflow component. (7) Logarithm of the column density of the first outflow component. (8) Central velocity of the second outflow component. (9) $b$ parameter of the second outflow component. (10) Logarithm of the column density of the second outflow component. We follow the convention of negative signs for outflow velocities and positive signs for inflow velocities}. Kinematic components marked with an asterisk are those discarded due to low S/N detection or contamination by other systems (see Section \ref{sec:voigt}.)
\label{tab:voigtfit2}
\end{table*}
Figure \ref{fig:example_outflow_inflow} shows, as an example, the kinematic components for two targets, M95-YSC1 and NGC4656-YSC1 (the rest of the sample can be found in Appendix). 
The absorption profiles of \SiII$\,$ 1526 (in blue) tracing neutral gas and \SiIV$\,$ 1402 (in red) tracing ionized gas, are shown on a common velocity axis, together with the best-fit Voigt model. The individual model components are drawn as dashed line: the components centered at zero km~s$^{-1}$ model the stationary gas, the blue-shifted components model the outflows, the red-shifted components model the inflows.

\begin{figure*}
    \includegraphics[width=0.5\textwidth]{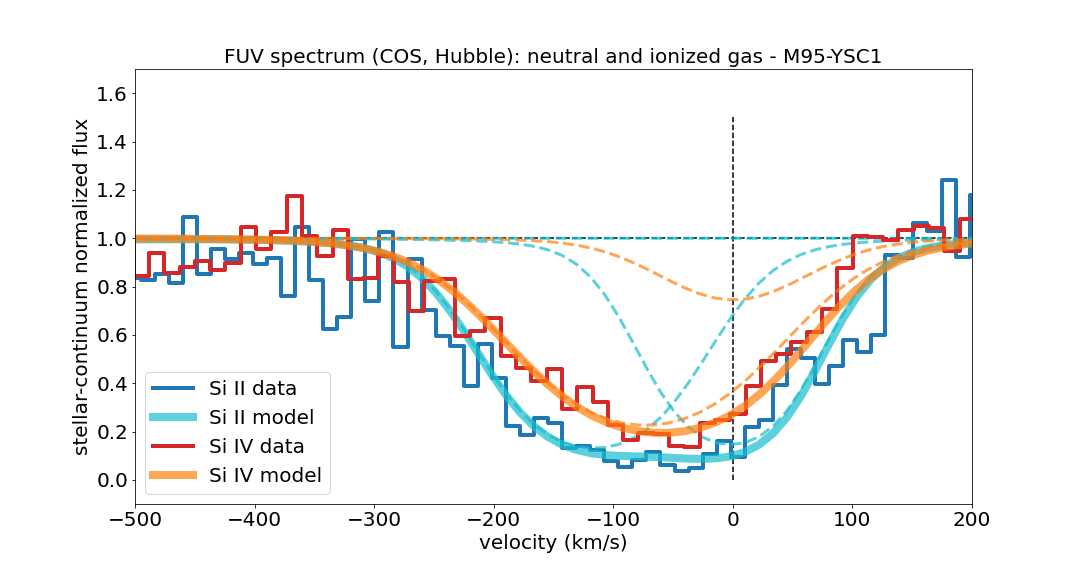}
    \includegraphics[width=0.5\textwidth]{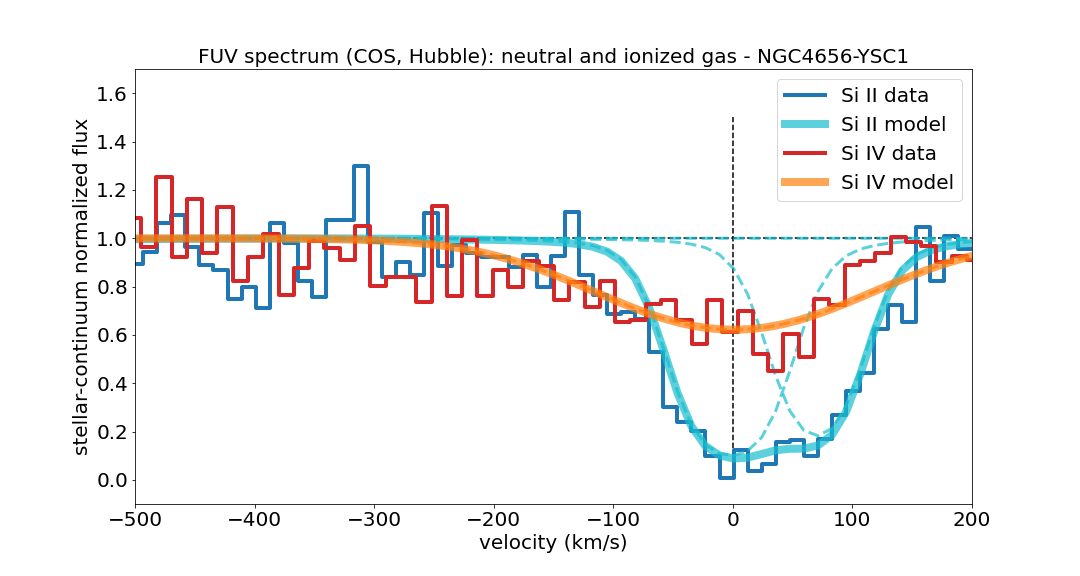}\\
     \caption{(Left) Kinematics of M95-YSC1 as an example of a target with detected outflow. The absorption profiles in the normalized spectrum for \SiII$\,$ 1526 (dark blue line) and \SiIV$\,$ 1402 (dark red line) are plotted on a common velocity axis, tracing neutral and ionized gas, respectively. The light blue and orange solid lines show the best-fit model, and the dashed color lines show the individual Voigt components of the best-fit model. A vertical black dashed line marks the zero velocity of the gas. (Right) Same as left panel but for NGC4656-YSC1, example of a target with a detected inflow.}
    \label{fig:example_outflow_inflow}
\end{figure*}

With our multi-component Voigt fit analysis, we detected outflowing gas in 16 out of 18 targets, either in the neutral or ionized phase, or both. We measured the significance of each outflowing gas component in terms of S/N. We estimated the noise by calculating the standard deviation in the stellar continuum over a spectral window without absorption or emission lines (typically 500-1000 km~s$^{-1}$ bluewards of the \SiII$\,$ 1526 and \SiIV$\,$ 1402 lines). The ratio between the absorbed flux (at its deepest absorption value in the model outflow component) and the noise was taken as a proxy of how significant the detection of the outflow is. We found significant detections at a level of up to 10$\sigma$, except for NGC1512-YSC2 \SiIV$\,$ and NGC1566-YSC1 \SiII$\,$ where S/N $<$ 2. We decided to set the bar for outflow detection at 2$\sigma$ and thus discard these two components. For M51-YSC2 we detected large outflow velocities compared to the targets of the sample with similar physical properties and stellar feedback. This cluster lies in the interacting-arm at the outskirt of NGC5194 (which here we refer to as M51), close to the companion galaxy NGC5195, and thus it is likely contaminated by large scale gas flows more than local feedback-induced kinematics. 
For this reason we discarded this fast outflow component of M51-YSC2 from the following analysis.

\begin{figure*}
    \centering
    \includegraphics[width=\textwidth]{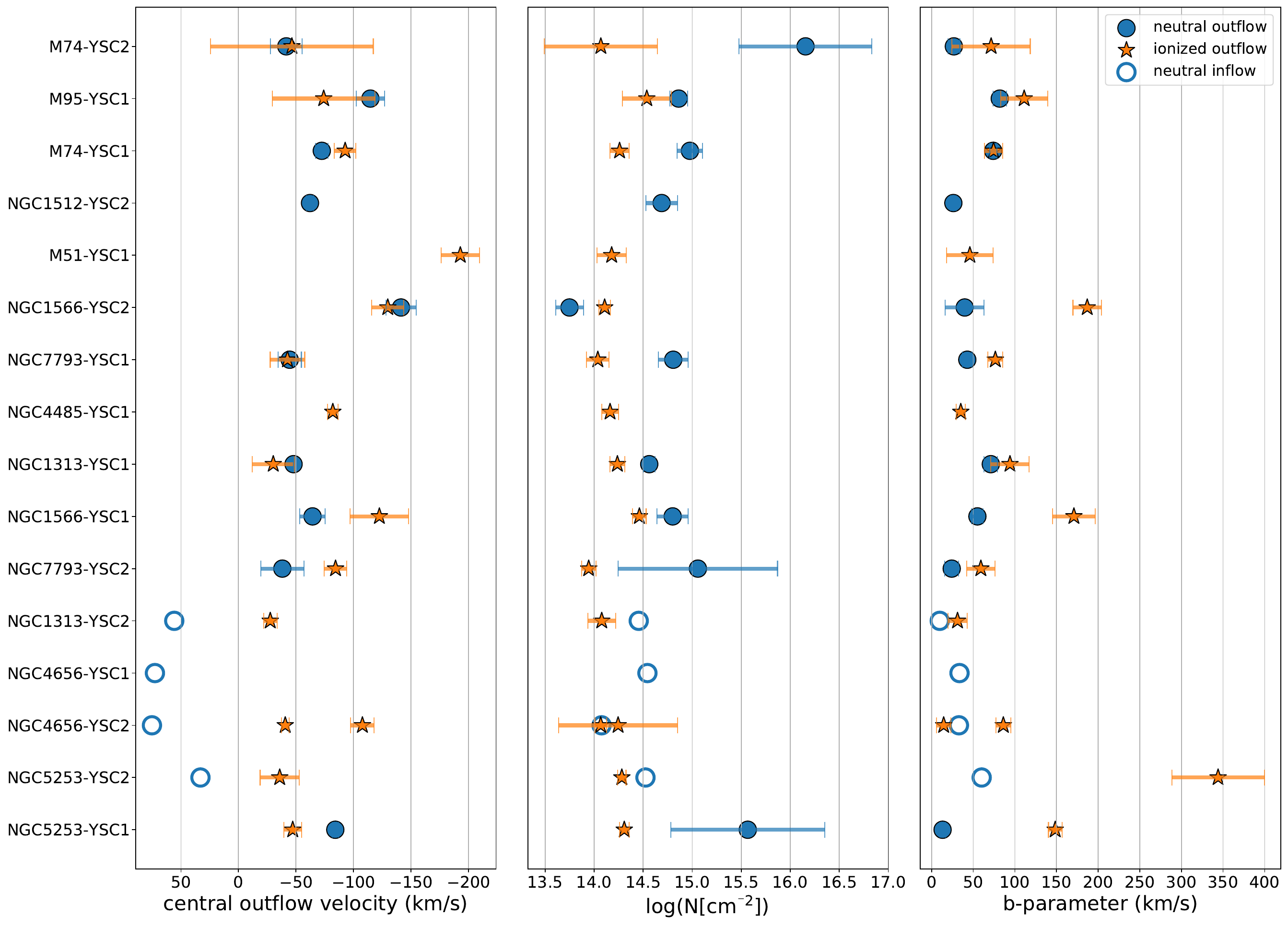}
     \caption{Best-fit values of the multi-component Voigt fit analysis of the interstellar gas kinematics. The panels from left to right show the central velocity, the column density of \SiII$\,$ and \SiIV$\,$ and the $b$ parameter of the outflow and inflow components detected). The blue circles represent the neutral gas traced by \SiII, the orange stars represent the ionized gas traced by \SiIV. The empty blue circles mark the gas components inflowing towards the target star cluster. M51-YSC1 is the exception target for which we used \CIV$\,$ instead of \SiIV. The targets are sorted by the age of the FUV dominant stellar population}.
    \label{fig:voigtfit}
\end{figure*}

Figure \ref{fig:voigtfit} shows the kinematics, column densities and $b$ parameters of the non-stationary gas, that is either outflowing or inflowing. The targets are sorted by increasing age of the FUV dominant stellar population, as the figures in Paper I.
The neutral gas kinematics is more reliable since it is traced by several transitions detected in our data (\AlII$\,$ 1670, \SII$\,$ 1250,1253,1259, \SiII$\,$ 1190,1193,1260,1526) rather than just a couple of lines (\SiIV$\,$ 1393,1402) as in the case of the ionized gas. This is likely the reason why the central velocity and $b$ parameter of the ionized gas components have larger errors than the neutral gas. The central velocity of the outflows is for most targets between 50 km~s$^{-1}$ and 150 km~s$^{-1}$. Only in a few targets, the two phases (neutral and ionized) of the outflows have the same velocity, while in the rest of the sample they differ but not systematically (see Figure \ref{fig:voigtfit}). We found a mean \SiII$\,$ column density of log$(N/\rm cm^{-2}) = 14.8 \pm 0.6 $ and a mean \SiIV$\,$ column density of log$(N/\rm cm^{-2}) = 14.2 \pm 0.2$ (where the errors represent the standard deviations). The $b$ parameter is for most targets between 10 km~s$^{-1}$ and 100 km~s$^{-1}$ with a few gas components having broad absorption profiles up to 350 km~s$^{-1}$. These extreme values of the $b$ parameter indicate a high turbulence in the gas examined. In fact, without turbulence the gas temperature would have to be higher than $T_{max}=2\times10^5$ K, which is the inferred maximum gas temperature given the detection of \NV$\,$ absorption lines and the absence of O V \citep{Tumlinson2017}. Using Equation \ref{eq:turbulence}, taken from \cite{Krogager2018}, we calculated that for the gas components with high values of the $b$ parameter, the turbulence energy actually dominates over the thermal energy:

\begin{equation}
    E_{turb} = 1/2 m_H b^2 - k_B T_{max}
    \label{eq:turbulence}
\end{equation}

where $m_H$ is the hydrogen atom mass and $k_B$ is the Boltzmann constant.
Another possibility is that the broad absorption is the result of many clouds with a smooth velocity variation tracing the bulk motion of the gas. In this scenario, a large number of clouds with a continuous distribution of velocities is required. However, we do not have the means and data resolution to discern which scenario is more likely.

\subsection{Gas covering fraction of the UV continuum background source}
\label{sec:covering}
Another important effect to be considered when studying gas properties using absorption lines, is the gas partial coverage of the background continuum source. In fact, both the column density $N_{ion}$ of a specific ion and the covering fraction $f_c$ contribute to the level of the absorbed continuum flux, according to the equation:

\begin{equation}
    F_{obs} = F_{cont} [1 - f_c (1 - e^{-\tau(N_{ion})})]
    \label{eq:covering}
\end{equation}

where $F_{obs}$ is the observed flux, $F_{cont}$ the continuum flux, $\tau$ the optical depth, which is proportional to the column density and depends on the oscillator strength of the specific ion transition. This implies that if the covering fraction is less than 1, our measurement of the column density is only a lower limit, since it would be higher if the same amount of flux were absorbed by a gas configuration that covers the background source only partially. We used four different transitions of the same ion (\SiII$\,$ 1190,1193,1260,1526) 
with different oscillator strengths to solve for column density and covering fraction simultaneously and break their degeneracy. The analysis has been performed as a function of velocity. We report the highest value of covering fraction as representative of the gas producing the deepest flux absorption. In principle, it is also possible to repeat the same measurement for the \SiIV, however due to the fact that we detected only two transitions of this ion, the values obtained are highly uncertain and are not reported. Table \ref{tab:covering} lists the covering fraction $f_c$ of the gas traced by \SiII, i.e. neutral gas, as well as the size in pc of the UV continuum source as measured by the half-light radius in the COS/NUV acquisition image.

\begin{table}
\centering
\begin{tabular}{llrrr}
target name & $f^{SiII}_c$ & $r_{UV}$ & log($N_H$) \\
 & & (pc) & \\ 
\hline
 M74-YSC2     & 1.00$^{+0.00}_{-0.08}$ &  9 & 20.0$\pm$0.2  \\
 M95-YSC1     & 1.00$^{+0.00}_{-0.00}$ & 12 & 19.7$\pm$0.1 \\
 M74-YSC1     & 0.97$^{+0.03}_{-0.03}$ & 13 & 20.0$\pm$0.2 \\
 NGC1512-YSC2 & 0.82$^{+0.01}_{-0.01}$ & 13 & 19.5$\pm$0.1 \\
 M51-YSC1     & 0.99$^{+0.01}_{-0.04}$ &  7 & 19.5$\pm$0.1 \\
 NGC1566-YSC2 & - & - & 19.73$\pm$0.04\\
 NGC1512-YSC1 & 0.87$^{+0.08}_{-0.08}$ & 24 & 20.22$\pm$0.04\\
 NGC7793-YSC1* & 0.96$^{+0.03}_{-0.03}$ &  3 & 20.6$\pm$0.1 \\
 NGC4485-YSC2 & 0.99$^{+0.01}_{-0.02}$ &  6 & 21.52$\pm$0.01 \\
 NGC4485-YSC1 & 0.98$^{+0.01}_{-0.03}$ & 13 & 20.78$\pm$0.03 \\
 NGC1313-YSC1* & 0.97$^{+0.01}_{-0.02}$ & 12 & 20.3$\pm$0.2 \\
 NGC1566-YSC1 & 0.92$^{+0.04}_{-0.02}$ & 57 & 20.33$\pm$0.03 \\
 NGC7793-YSC2* & 0.84$^{+0.01}_{-0.01}$ & 11 & 20.6$\pm$0.1 \\
 NGC1313-YSC2* & 0.95$^{+0.02}_{-0.02}$ &  8 & 21.57$\pm$0.01 \\
 NGC4656-YSC1 & 0.99$^{+0.01}_{-0.02}$ &  8 & 21.79$\pm$0.01 \\
 M51-YSC2     & 0.98$^{+0.02}_{-0.06}$ &  5 & 20.55$\pm$0.07 \\
 NGC4656-YSC2 & 0.99$^{+0.01}_{-0.01}$ &  4 & 21.79$\pm$0.01 \\
 NGC5253-YSC2* & 0.80$^{+0.00}_{-0.00}$ & 10 & 20.71$\pm$0.09 \\
 NGC5253-YSC1* & 0.91$^{+0.00}_{-0.00}$ & 10 & 21.2$\pm$0.1 \\
\hline
\end{tabular}
\caption{Gas covering fraction based on the \SiII$\,$ tracer listed for each target, NUV half-light radius of the continuum source, total hydrogen column density from \HI$\,$ absorption. NGC1566-YSC2 has been excluded from the covering fraction analysis due to an insufficient number of \SiII$\,$ transitions detected. The targets with their name followed by an asterisk are those that are contaminated by the Magellanic System absorption and therefore might have the values of column density overestimated.}
\label{tab:covering}
\end{table}

All targets have a covering fraction of neutral gas between 0.8 and 1 with the majority being above 0.95. We conclude that the column density, which is measured under the assumption of unity covering fraction and reported in Table \ref{tab:voigtfit}, is not significantly affected by the effect of partial coverage. 

\section{Interstellar gas outflows} \label{sec:outflows}

\subsection{Correlation between outflow velocities and stellar properties}
\label{sec:out_correlations}

We investigated the presence of any correlation between the physical properties of the stellar population of our targets and the outflow velocities of the ISM around them. We looked for the presence of any correlation separately for the neutral and the ionized outflowing gas. The stellar population properties that we tested are: ages (of the young/old population, of the FUV dominant/fainter population, mass-weighted age), mass (of the young/old population, of the FUV dominant/fainter population, total mass), mechanical energy integrated since the formation of the stars (stellar winds, SN explosions, total energy), photon production rate, instantaneous mechanical luminosity (stellar winds, SN explosions, total power). We used the Kendall rank correlation coefficient $\tau$\footnote{We note that rank correlation coefficients do not take into account the uncertainties of the measurements.} to test the significance of the correlation between the outflow velocities and the different parameters listed above. The Kendall $\tau$ is a statistics used to measure the ordinal association between two quantities and therefore is sensitive not only to linear correlation but more generally to any monotonic function. For determining the significance of any correlation indicated by the Kendall $\tau$, we calculated the p-value for a hypothesis test whose null hypothesis is an absence of association ($\tau$=0). The p-value can be seen as the probability of obtaining the same level of association after randomizing the order of the two sets of quantities on the axis. We note that $\tau$ and the p value might not be very robust to outliers.

\begin{figure*}
    \centering
    \includegraphics[width=\textwidth]{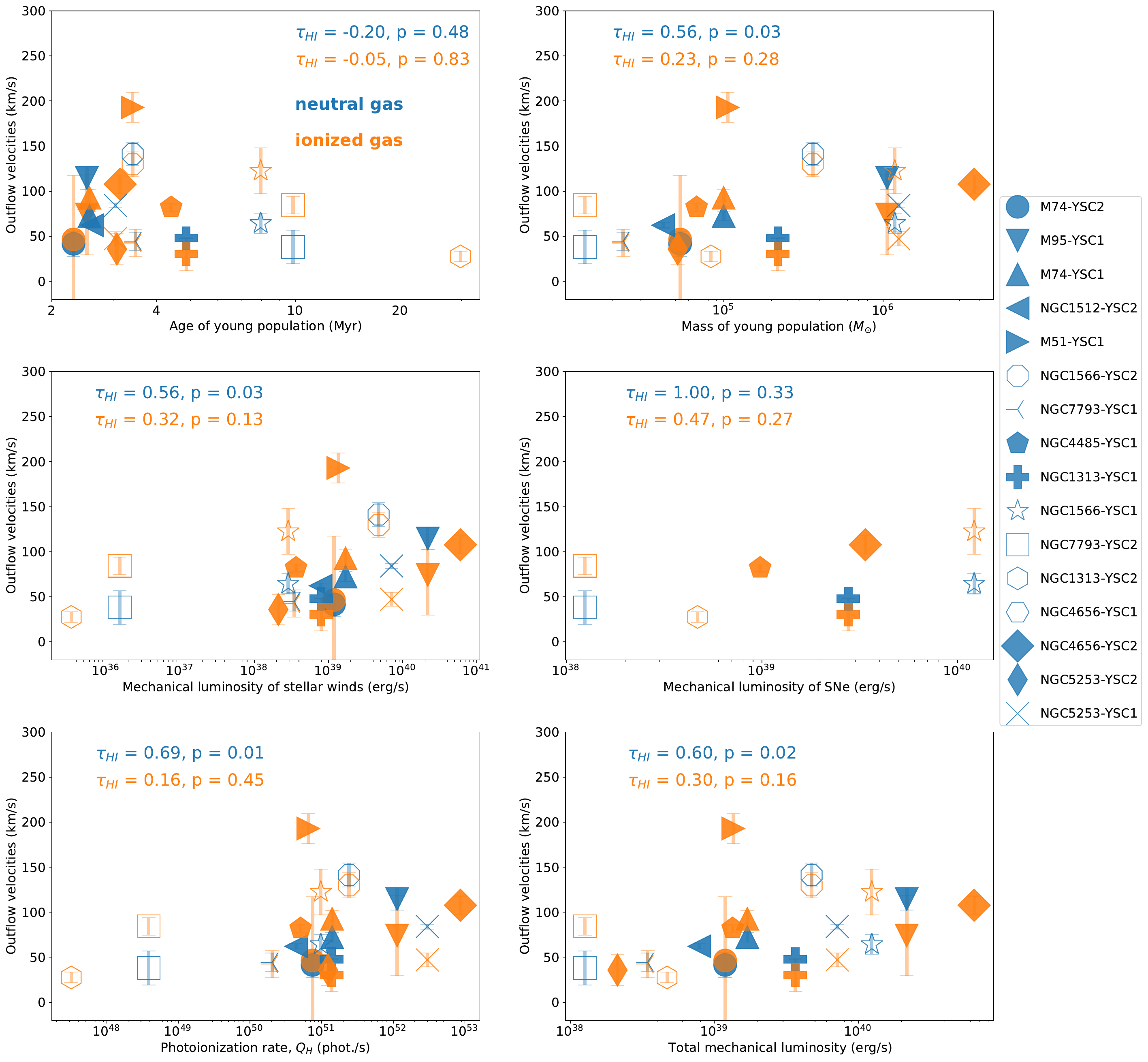}
     \caption{Outflow velocities plotted against six different stellar properties: age of young stellar population, mass of the young stellar population, mechanical luminosity of stellar winds, mechanical luminosity of SNe, photon production rate and total mechanical luminosity from stellar winds and SNe. The blue data points represent the neutral gas, whereas the orange data points represent the ionized gas. Hollow markers (NGC1566-YSC2, NGC4485-YSC2, NGC1566-YSC1, NGC7793-YSC2, NGC1313-YSC2, NGC4656-YSC1) indicate the targets with an age uncertainty larger than 25\%, which gives a sense of which data points are the most uncertain in all the plots. Error bars for the outflow velocities are plotted. $\tau$ and p are two parameters of the Kendall statistics performed to assess the presence or lack of correlation (see text).}
    \label{fig:correlations}
\end{figure*}

Figure \ref{fig:correlations} shows a plot of the outflow velocities (neutral phase in blue, ionized phase in orange) against six different stellar properties among the ones mentioned above: age of young stellar population, mass of the young stellar population, mechanical luminosity of stellar winds, mechanical luminosity of SNe, photon production rate of Hydrogen and total mechanical luminosity from stellar winds and SNe \footnote{For the photon production rate, the mechanical energy and the mechanical luminosity, in this figure we sum together the contributions of both stellar populations of the clusters that have two populations.}. The error bars represent the uncertainties of the velocities. For the target NGC4656-YSC2, which is the only one featuring two distinct outflowing gas components, we used the faster component of the two detected for the study of correlations. The purpose of Figure \ref{fig:correlations} is to visualize the presence of correlation or lack thereof between outflow velocity and the main stellar properties. The hollow data points mark those targets whose age errors are larger than 25 \%. Each panel shows the Kendall $\tau$ and p-value for both phases of the gas. Based on these coefficients, we found a significant correlation (p-value lower than 3\%) between the neutral outflow velocities and the following stellar properties: mass of the young stellar population, photon production, mechanical luminosity of stellar wind and total mechanical luminosity from stellar winds and SNe. The latter two properties show a correlation also with the outflow velocities of the ionized gas, although not very significant (p-values of 13\% and 16\%). Because M51-YSC1 is an outlier in the correlation between total mechanical luminosity and ionized outflow velocity, we computed the Kendall $\tau$ without this target and obtained values indicating a more significant correlation, $\tau=0.36$ and p-value of 10\%. The data points feature a significant degree of scatter, which can be due to the simultaneous dependence of the outflow velocities on multiple parameters. The scatter might also be associated with the uncertainty in the values of the stellar properties (see Paper I). Moreover, we note that the observed correlations are not independent, since the mechanical power and the photon production rate are proportional to the mass of the stellar population.

\begin{figure*}
    \centering
    \includegraphics[width=0.9\linewidth]{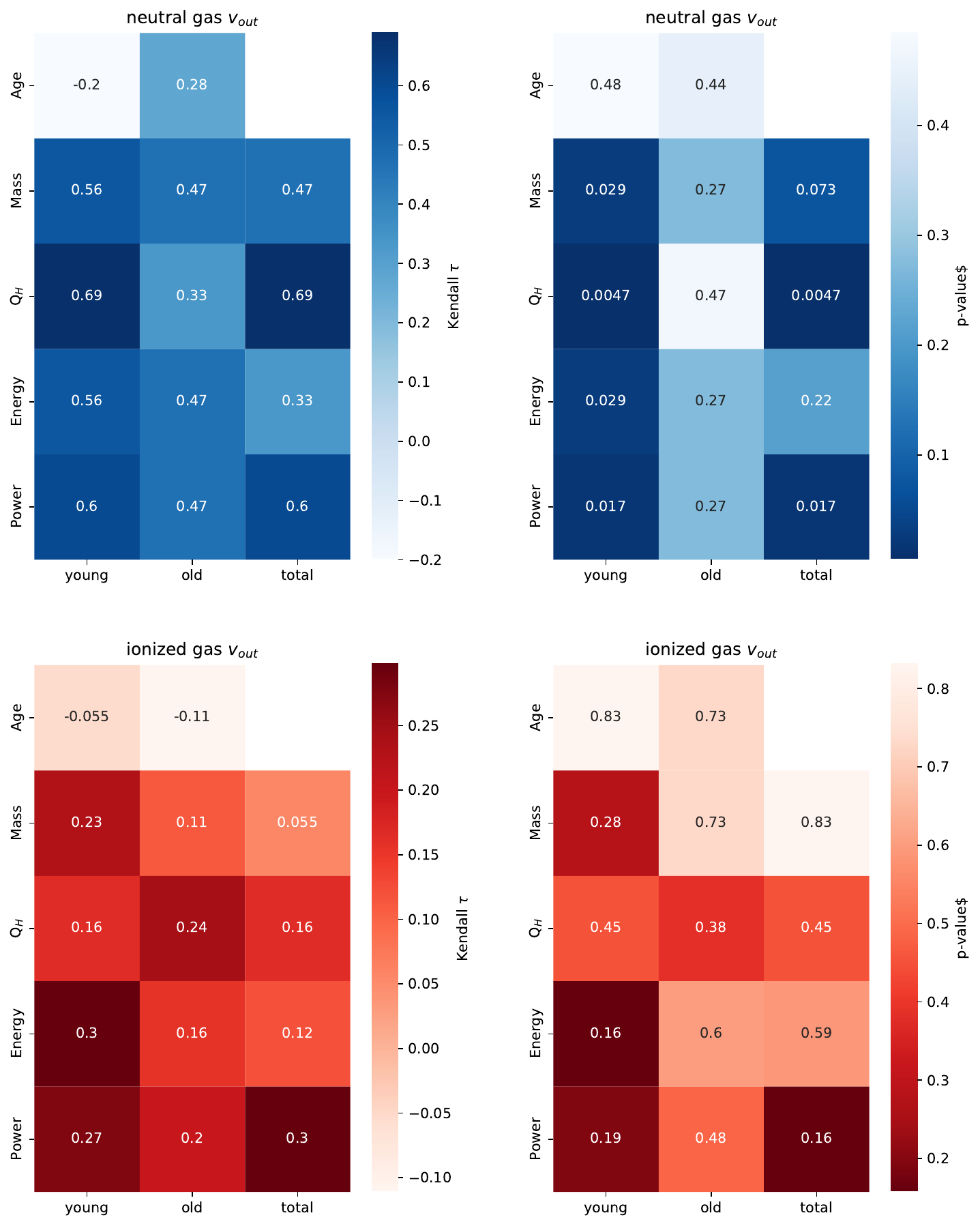}
    \caption{Kendall statistics for the correlation of outflow velocities with different stellar properties (age, mass, photon production rate, mechanical energy of stellar winds and SNe, mechanical luminosity, here labelled ``power", of stellar winds and SNe) separated into three contributions: young stellar population, old stellar population, total stellar population. The blue heat maps show the correlations for the neutral outflowing gas, whereas the red heat maps show the correlations for the ionized outflowing gas. The left panels report the Kendall $\tau$ coefficients while the right panels report the p-values with the reversed color-bar, so that small values, meaning more significant correlations, appear darker.}
    \label{fig:heatmap}
\end{figure*}

Figure \ref{fig:heatmap} shows a heat map to summarize and compare the stellar properties with outflow velocities to assess if any of them correlate with each other.
The correlation between the mass of the stellar population and the outflow velocity indicates that for larger masses the stellar feedback increases, meaning a larger amount of energy and momentum is generated, which drives faster outflows.
Additionally, we found the photon production rate $Q_H$ to correlate with the outflow velocities. 
It is worth noting that the correlation is generally better for the young population than the old or combined populations. This may be taken as a sign that the old population is in the background/foreground and possibly not contributing to the superbubble outflow. Alternatively, it could be because the age estimates are less precise for the older stellar populations, which are typically fainter. Finally, some models predict an anti-correlation between the cluster age and the outflow velocity for cluster winds \citep{Silich2013}, which would be consistent with the negative correlation coefficients we recovered. However, the high p-value indicates that this negative correlation is not significant.

\subsection{Outflows energetics} \label{sec:out_energetics}

In this section we report the estimated outflow energetics, namely the momentum flux, the kinetic energy of the outflowing gas and its kinetic power, i.e. the rate at which the kinetic energy is transported outward. For these estimates, it is essential to have a measurement of the mass of the outflowing gas. In order to measure the mass, we made the following assumptions: (i) we assumed the outflow to be a spherical shell of gas with uniform density; (ii) the outflow starts with a zero velocity when the central cluster forms and accelerates to the observed velocity with a constant acceleration equal to $a=v_{out}/t_{cluster}$, meaning that the integrated radius $R$ of the spherical shell is $R=v_{out}\,t_{cluster}/2$, where $v_{out}$ is the outflow velocity and $t_{cluster}$ is the age of the young stellar population; (iii) we converted Si mass to total gas mass using the measured oxygen abundance and solar abundance patterns.
There are 6 targets where only the high or low ions are detected in the outflow, 5 targets with overlapping velocity errors, and only 4 targets that have different high/low ion velocities. 
For consistency, we have measured for all the targets sizes and masses for the neutral and ionized outflows separately \footnote{The two outflows contain in fact all the gas phases, but we will refer to them as the neutral and the ionized outflows to indicate the line used for the detection.} However, for all the targets where both low and high ions are detected in the outflow, because of the significant velocity overlap (see Figure \ref{fig:voigtfit}), we assumed that all ions are tracing the same outflow, and we used the Si II tracer to calculate the mass and the energetics since Si IV might give more uncertain results. In Section \ref{sec:radius} we further discuss the assumptions and the uncertainties of the outflow radius measurement. 
For a target galaxy with an ISM where the Si abundance is the same as the solar abundance, the ratio between Si atoms and H atoms is $N_{Si}/N_{H}=3.23 \times 10^5$. We corrected this value using the solar oxygen abundance recommended by \cite{Asplund2009}:

\begin{equation}
    N_{Si}/N_{H}=3.23 \cdot 10^5 \times 10^{\rm log(O/H) - 8.69}
    \label{eq:Zscale}
\end{equation}

The values of $\rm log(O/H)$ are tabulated in Table 1 of Paper I.
The outflow mass is therefore given by the following equation:

\begin{equation}
    M_{out} = 4 \pi R^2 (N_{H}/N_{Si}) N_{Si_{out}} m_p\, \mu\, / \xi
    \label{eq:energy}
\end{equation}

where $N_{Si_{out}}$ is the column density of the outflow component in \SiII$\,$ for the neutral outflow or \SiIV$\,$ for the ionized outflow, $m_p$ is the mass of a proton, $\mu=1.4$ is the mean molecular weight (to account for the fact that there is a small percentage of heavier elements in the ISM in addition to hydrogen atoms). $\xi$ is the ionization fraction, i.e. the fraction of Si atoms in the ionization state of \SiII$\,$ (16 $\pm$ 4 \%) or \SiIV$\,$ (10 $\pm$ 2 \%). These values were derived in \cite{Chisholm2016} for a sample of large-scale winds. As a sanity check, we calculated the equivalent widths ratios of \SiIV$\,$ over \SiII$\,$ of the CLUES outflows, and they all agree with the interval 0.91$\pm$0.25 reported in \cite{Chisholm2016}.
For the target NGC4656-YSC2 that has multiple outflow components, we calculated the mass for each component using the respective Si column density, and then we summed them up. 

The momentum flux of the outflow is calculated as:
\begin{equation}
    \label{eq:pout}
    \dot{p}_{out} = \sum_i m_{out,i} v_{out,i} / \Delta t
\end{equation}
where $m_{out,i}$ and $v_{out,i}$ are the mass and the central velocity, respectively, of the outflow components, while $\Delta t$ is the timescale of the outflow, which we assume equal to the cluster age.
The kinetic energy of the outflow is given by:
\begin{equation}
    \label{eq:Kout}
    K_{out} = \sum_{i} \frac{1}{2}\, m_{out,i} v^2_{out,i}
\end{equation}
We remind the reader that for all targets except NGC4656-YSC2 the sum in Equations \ref{eq:pout} and \ref{eq:Kout} reduces to only one term. Similarly, the kinetic power (mechanical luminosity) of the outflow is given by:
\begin{equation}
    P_{out} = \sum_i \frac{1}{2}\, m_{out,i} v^2_{out,i} / \Delta t
\end{equation}

Table \ref{tab:energetics} reports the values of outflow velocities and energetics estimated for the CLUES targets. The size of the outflows are a few hundred pc, although these estimates carry large uncertainties and so do the quantities derived with them  (see Section \ref{sec:radius}). Considering all targets, the masses for both the neutral and ionized outflows range between $10^5$ and $10^6\, \msun$. The momentum flux spans between $10^{31}$ and $10^{32}$ dyne.
The kinetic energies of the outflows for most targets fall in the range $10^{51}$ and $10^{54}$ erg and the kinetic powers are between $10^{37}$ and $10^{40}$ erg/s. We discuss these values in Section \ref{sec:discussion}.

\begin{table*}
\centering
\begin{tabular}{lccccccccc}
\hline

target & $v_{out} \pm b$ & $v_{out,ion} \pm b$ & $R_{out}$ & $R_{out,ion}$ & $M_{out}$ & $M_{out,ion}$ & $\dot{p}_{out}$ & $K_{out}$ & $P_{out}$ \\

 & [km~s$^{-1}$] & [km~s$^{-1}$] & [pc] & [pc] & [\msun] & [\msun] & [dyne] & [erg] & [erg/s] \\
 (1) & (2) & (3) & (4) & (5) & (6) & (7) & (8) & (9) & (1)\\
\hline
 M74-YSC2     & 42$\, \pm \, $27  & 46$\,\pm \, $72   &  49 &  55 & $9.3 \cdot 10^{5}$ & $1.5 \cdot 10^{4}$ & $1.1 \cdot 10^{32}$ & $1.6 \cdot 10^{52}$ & $2.2 \cdot 10^{38}$ \\
 M95-YSC1     & 115$\, \pm \, $82 & 74$\,\pm \, $111  & 147 &  95 & $1.4 \cdot 10^{5}$ & $4.3 \cdot 10^{4}$ & $3.9 \cdot 10^{31}$ & $1.8 \cdot 10^{52}$ & $2.2 \cdot 10^{38}$ \\
 M74-YSC1     & 72$\, \pm \, $74  & 93$\,\pm \, $74   &  95 & 121 & $2.3 \cdot 10^{5}$ & $1.2 \cdot 10^{5}$ & $4.1 \cdot 10^{31}$ & $1.2 \cdot 10^{52}$ & $1.5 \cdot 10^{38}$ \\
 NGC1512-YSC2 & 62$\, \pm \, $26  & -                 &  83 & -& $7.2 \cdot 10^{4}$ & -                  & $1.1 \cdot 10^{31}$ & $2.8 \cdot 10^{51}$ & $3.4 \cdot 10^{37}$ \\
 M51-YSC1     & -                 & 193$\,\pm \, $46  & -& 335 & -                  & $5.8 \cdot 10^{5}$ & $2.1 \cdot 10^{32}$ & $2.2 \cdot 10^{53}$ & $2.0 \cdot 10^{39}$ \\
 NGC1566-YSC2 & 141$\, \pm \, $40 & 130$\,\pm \, $187 & 245 & 226 & $3.7 \cdot 10^{4}$ & $1.1 \cdot 10^{5}$ & $9.6 \cdot 10^{30}$ & $7.3 \cdot 10^{51}$ & $6.8 \cdot 10^{37}$ \\
 NGC7793-YSC1 & 45$\, \pm \, $43  & 43$\,\pm \, $77   &  79 &  75 & $1.4 \cdot 10^{5}$ & $3.4 \cdot 10^{4}$ & $1.1 \cdot 10^{31}$ & $2.7 \cdot 10^{51}$ & $2.5 \cdot 10^{37}$ \\
 NGC4485-YSC1 & -                 & 82$\,\pm \, $35   & -& 184 & -                  & $2.2 \cdot 10^{5}$ & $2.6 \cdot 10^{31}$ & $1.5 \cdot 10^{52}$ & $1.1 \cdot 10^{38}$ \\
 NGC1313-YSC1 & 48$\, \pm \, $71  & 30$\,\pm \, $94   & 119 &  75 & $2.8 \cdot 10^{5}$ & $8.4 \cdot 10^{4}$ & $1.7 \cdot 10^{31}$ & $6.4 \cdot 10^{51}$ & $4.2 \cdot 10^{37}$ \\
 NGC1566-YSC1 & 64$\, \pm \, $55  & 122$\,\pm \, $171 & 261 & 496 & $4.7 \cdot 10^{5}$ & $1.2 \cdot 10^{6}$ & $2.4 \cdot 10^{31}$ & $1.9 \cdot 10^{52}$ & $7.7 \cdot 10^{37}$ \\
 NGC7793-YSC2 & 38$\, \pm \, $24  & 84$\,\pm \, $59   & 191 & 423 & $1.4 \cdot 10^{6}$ & $8.6 \cdot 10^{5}$ & $3.5 \cdot 10^{31}$ & $2.1 \cdot 10^{52}$ & $6.7 \cdot 10^{37}$ \\
 NGC1313-YSC2 & -                 & 28$\,\pm \, $31   & -& 419 & -                  & $1.8 \cdot 10^{6}$ & $1.1 \cdot 10^{31}$ & $1.4 \cdot 10^{52}$ & $1.5 \cdot 10^{37}$ \\
 NGC4656-YSC1 & -                 & -                 & -& -& -                  & -                  & -                    & -                   & -                   \\
 NGC4656-YSC2 & -                 & 108$\,\pm \, $86  & -& 172 & -                  & $6.0 \cdot 10^{5}$ & $2.1 \cdot 10^{32}$ & $8.4 \cdot 10^{52}$ & $8.6 \cdot 10^{38}$ \\
 NGC5253-YSC2 & -                 & 36$\,\pm \, $344  & -&  56 & -                  & $8.3 \cdot 10^{4}$ & $6.2 \cdot 10^{30}$ & $1.1 \cdot 10^{51}$ & $1.1 \cdot 10^{37}$ \\
 NGC5253-YSC1 & 84$\, \pm \, $13  & 47$\,\pm \, $148  & 130 &  73 & $5.4 \cdot 10^{6}$ & $1.5 \cdot 10^{5}$ & $9.5 \cdot 10^{32}$ & $3.8 \cdot 10^{53}$ & $4.0 \cdot 10^{39}$ \\
\hline
\end{tabular}
\caption{Outflows energetics. (1) target name. (2) velocity range of the neutral outflow. (3) velocity range of the ionized outflow. (4) Neutral outflow radius. (5) Ionized outflow radius (6) Mass of the neutral outflow. (7) Mass of the ionized outflow. (8) Momentum flux of the outflow. (9) Kinetic energy of the outflow. (10) Kinetic power of the outflow. The uncertainties on the radii are at least 50\%, and they represent the main source of errors for the outflow masses and energetics. Therefore, the values of the energetics in the table also have similar errors or larger, due to the necessary assumptions (see Section \ref{sec:discussion}).}
\label{tab:energetics}
\end{table*}

\section{Discussion} \label{sec:discussion}

We quantified the stellar feedback of YSCs and measured the kinematics of the interstellar gas around the stellar populations, using FUV spectroscopy.

\subsection{Uncertainties on the derived stellar physical parameters and outflow properties}

This work relies importantly on the results obtained in Paper I, where we modeled the FUV stellar continuum with stellar population models based on Starburst99 libraries. 
The detection of P-Cygni lines in the FUV allowed us to constrain the age of the young stellar population in a robust way, as confirmed by comparing different methods (Paper I). However,
there might be systematic issues associated with the models because our theoretical understanding of massive stars is still incomplete \citep{Eldridge2022}. The IMF could be another issue, since even a single very massive star can have an impact on the integrated light and mass loss rates \citep[e.g.][]{Martins2022}. Moreover, the uncertainty in the age estimates also propagates into the stellar feedback quantities.
The uncertainty in the stellar properties likely contributes to the scatter observed in the correlation plots of Figure \ref{fig:correlations}.

The second part of this work consists in deriving the kinematics of the interstellar gas by modeling the FUV absorption lines. The FUV absorption profile is modeled after normalizing the FUV spectra by the best-fit stellar continuum. We mentioned in Section \ref{sec:kinematics} that for five targets we improved locally the model of the stellar continuum for the wavelength range around a few P-Cygni lines (e.g. \SiIV, \NV). 
P-Cygni lines are the signature of stellar winds, which contribute in absorbing the stellar flux and therefore must be modeled as accurately as possible to disentangle the absorption due to the ISM gas. Despite the normalization of the FUV spectra introducing a source of error, we detected several absorption lines with an S/N up to 10 (as described in Section \ref{sec:kinematics}). In addition, we stress that thanks to the availability of multiple transitions of the same ion or different ions with a similar ionization potential, we are able to constrain the velocity structure of the multiple gas components, by tying the velocity structure separately for the low ionization and high ionization lines.  

We can distinguish three different types of motion for the gas particles: outflow bulk motion (i.e. what we call outflow velocity, represented by the central velocity of the kinematic component), thermal random motion and turbulent chaotic motion. The kinematic components that we used to model the absorption profiles are typically broad, with Voigt $b$ parameters that in a few cases can exceed 100 km~s$^{-1}$. In order to explain such large values, we must invoke a very turbulent gas where the energy due to the turbulent motion dominates over the thermal energy (see Section \ref{sec:voigt_results}). 

\subsection{Stellar feedback as a driver of local outflows}
\label{sec:driving}

The correlation analysis between the outflow velocities and the physical properties of the star clusters sheds light on the role played by the latter in driving local outflows. Although the trends that we found in Figure \ref{fig:correlations} are for only a small number of targets and more data points would be needed to make our results more robust, we discuss here the implications of the significant correlations found according to the Kendall statistics.

The correlation of the outflow velocities with the photon production rate suggests that the direct and thermal radiation pressure exerted by the ionizing photon flux and the ionized gas pressure are an important component of the feedback. This is consistent with the result of \cite{Lopez2014,McLeod2021,Barnes2021,Chevance2022,DellaBruna2022}, who found that the thermal radiation pressure dominates the pre-SN feedback and drives the \HII$\,$ expansion. Moreover, we found a correlation between the outflow velocity and the mechanical luminosity of stellar winds as well as that of stellar winds and SNe combined. The fact that both the photon production rate and the mechanical luminosity of stellar winds correlate with the outflow velocities indicates that the pre-SN feedback is important and sufficient to drive outflows alone. This is consistent with numerical simulations that highlight the fundamental role that pre-SN feedback has in re-processing the gas of the natal cloud where SN will explode \citep{Bending2022}.

We found a weak correlation with the total (young and old stellar population together) mechanical energy released by stellar winds and/or SNe (see Figure \ref{fig:heatmap}). However, when the young stellar population is isolated, the correlation between mechanical energy and outflow velocity becomes more significant, as much as the one for mechanical luminosity. This is consistent with the old stars being in the background rather than in the cluster, and therefore not driving the observed mechanical energy of the outflows.
One stellar property that does not correlate with the outflow velocities is the age of the stellar populations. However, the age dependence is challenging to study given that several other parameters contribute to determining the outflow velocity and the CLUES sample is not large enough to control for these parameters (e.g. mass of the clusters, mass of the outflowing gas, morphology of the ISM). 


\subsection{Challenges in the measurement of the outflow radius}
\label{sec:radius}
The major source of uncertainty when estimating the outflow mass, and therefore the outflow energetics, is the measurement of the outflow radius. 
Absorption lines do not provide information on distance unless various models or assumptions are adopted, and the absorbing gas is often too diffuse to trace in emission. 
For this reason, it is particularly challenging to obtain a measurement of the distance between the continuum source and the site where most of the absorption occurs. It is also crucial to distinguish other gas systems that contaminate the observed absorption profiles, e.g. Milky Way (most targets), Magellanic System (NGC1313, NGC7793, NGC5253) and galaxy companion interactions (M51).

In our analysis, we assume the outflow to be a uniformly accelerating expansion that started when the star cluster formed. Its current radius depends on the outflow velocity and the cluster age. For the latter, we used the age of the young population, assuming that it drives the expansion of the bubble (see Table \ref{tab:energetics}). 
However, the assumption of a constant outflow acceleration is likely too simplistic since the outflow is likely to have a less uniform motion depending on the driving feedback and the interaction with the interstellar gas.

Previous works in the literature either assume a fiducial value for the outflow radius \citep[e.g. 1-5 kpc for galactic outflows,][]{Rupke2005,Martin2012}, or use a photoionization model that fit the observed column densities \citep{Chisholm2016,Chisholm2017} or relate it to the starburst radius \citep[e.g.][]{Heckman2015,Xu2022}. 
\cite{Hayes2023} introduced a novel method to estimate the radius of a galactic outflow as the integral of its velocity over the starburst lifetime, based on the fact that the outflowing gas accelerates under the effect of the stellar feedback and therefore its velocity monotonically increases with time. However, the lack of a clear correlation between v$_{out}$ and age in our sample does not allow us to apply a similar method in this work. The radii derived here are larger than the inner radii determined for the large-scale outflows in \cite{Chisholm2016,Chisholm2017} using photoionization models, while they overlap with the smaller sizes reported for the CLASSY sample \citep{Xu2023} using the non-resonant absorption lines. 

The assumption of a spherical outflowing shell makes the calculations notably easier but holds only in the scenario in which the density of the ISM is uniform. In reality, the gas density decreases with increasing distance from the galactic plane in a disk galaxy, like most of our targets. Therefore, the interstellar gas distribution is stratified, much like in a hydrostatic atmosphere, with a scale height of 100 pc for a typical galaxy \citep{Dyson1997}. Our estimated radii for the gas outflows are of the same approximate size. In such cases, a more realistic shape for the outflowing shells is a bubble elongated in the direction perpendicular to the galaxy disk, as the lower density gas offers the lower resistance path for the outflow. In such scenario, it is possible that a portion of the outflowing gas reaches the circumgalactic medium polluting it with metals and possibly falls back to the disk in a fountain type of flow, as shown in simulations \citep[e.g.][]{Kim2018}. We note that all the targeted spiral galaxies are observed face-on, and hence the line of sight to the star clusters coincides with the direction of elongation of the outflowing shell. 

\subsection{Comparison between feedback and outflow energetics}

\begin{figure}
    \centering
    \includegraphics[width=\columnwidth]{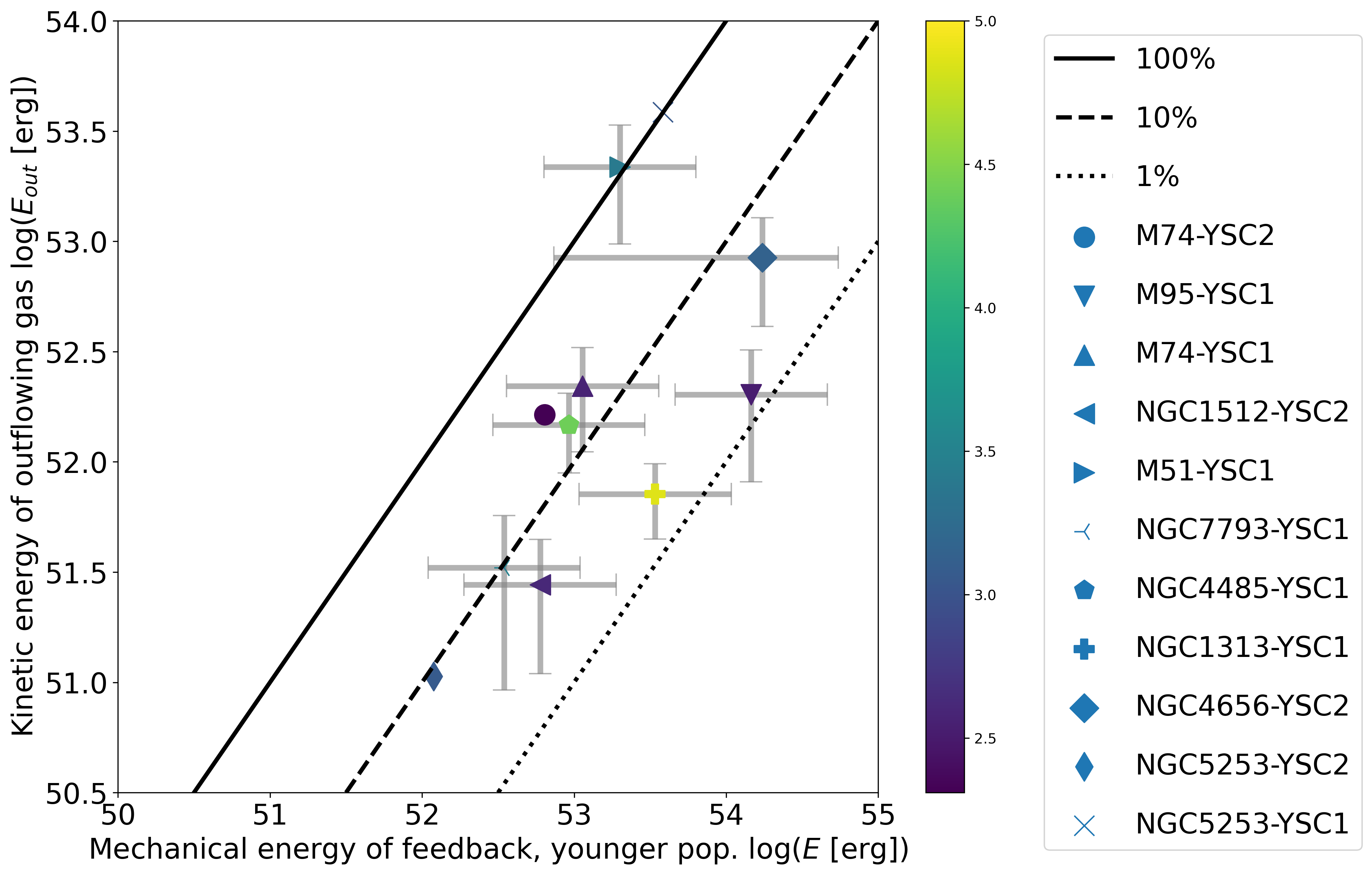}
    \includegraphics[width=\columnwidth]{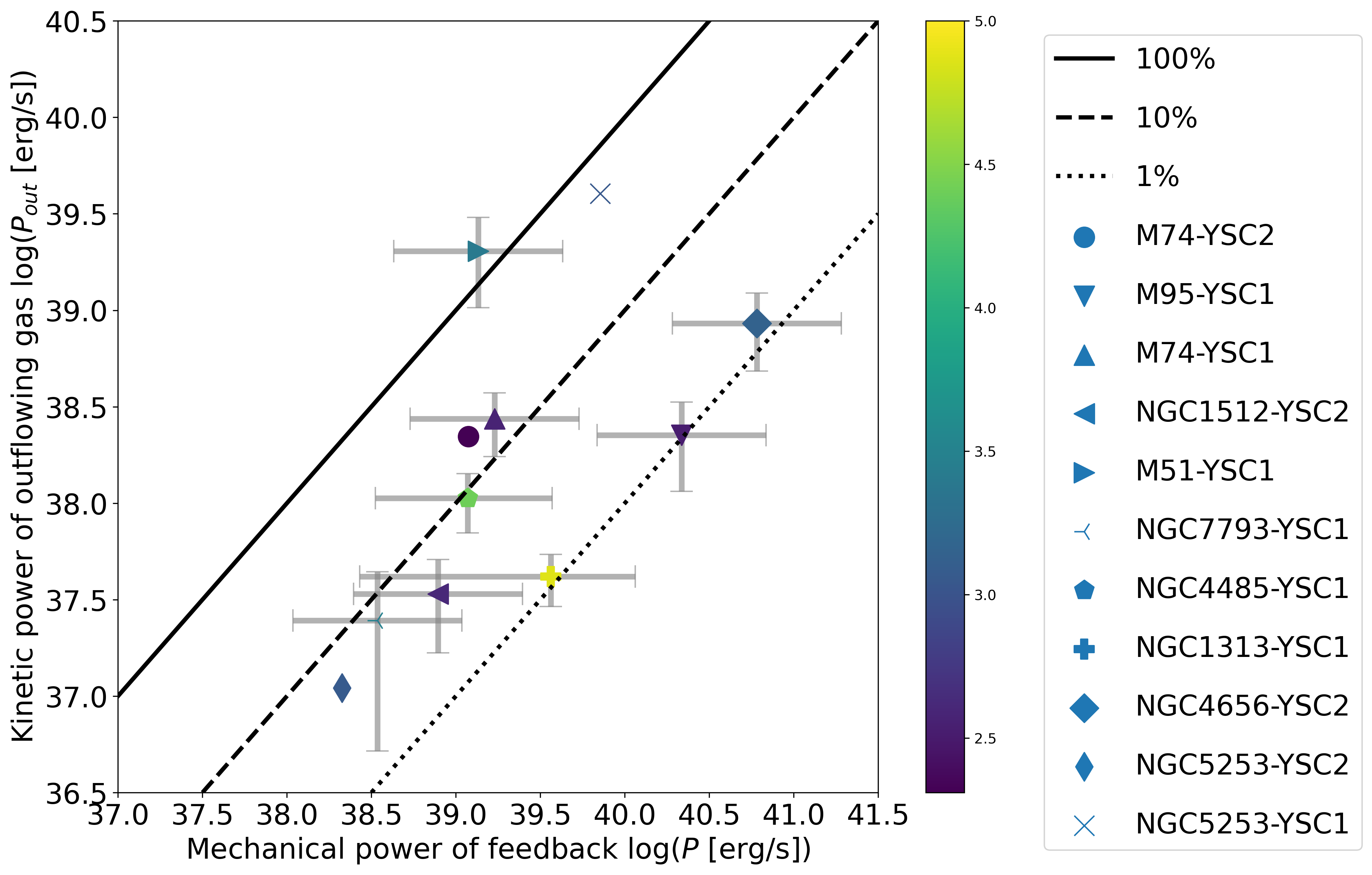}
    \caption{Comparison between stellar feedback and outflow mechanical energy and mechanical power. Only the clusters with an age error smaller than 25\% are shown. The markers are color coded by the age of the cluster (young stellar population), expressed in Myr on the colorbar. The solid line, dashed line and dotted line show the 1:1, 1:10 and 1:100 relations respectively. For the data points with no error bars, the uncertainties could not be determined realistically due to numerical precision issues.}
    \label{fig:compare}
\end{figure}

We compared two outflow properties, namely the kinetic energy and the kinetic power, with the respective physical quantities of the stellar feedback, considering only the contribution of the young stellar population. Figure \ref{fig:compare} shows such comparison for the CLUES clusters whose age has been determined most accurately (i.e. we discarded clusters with age errors larger than 25\%).
The markers are color-coded by the age of the cluster (the young stellar population).
The mechanical energy and mechanical luminosity comparisons show that the coupling efficiency of the feedback with the interstellar gas ranges between a few percent up to 100 percent.

We calculated the ratio between mechanical energy of the feedback and that of the outflow, finding a mean value of 10$\pm$8 \%. The large error is determined by the standard deviation and simply reflect the large scatter recovered for the coupling efficiency. Similarly, we calculated the ratio between the mechanical power measured in the outflow and the one imparted by the stellar population, finding a mean of 7$\pm$6 \%. These ratios represent the integrated and instantaneous coupling efficiency between the stellar feedback and the interstellar gas in the phases that are traced by the FUV lines used in this work. The values found in this work agree with those reported for galactic-scale outflows \citep[1 to 20 \%,][]{Chisholm2017} as well as those reported by \cite{Chevance2022}, who found a range of 8 to 50 \% for pre-SN feedback within star-forming regions in local spiral galaxies. These results indicate that a substantial fraction of the energy provided by the stellar feedback is dissipated, e.g. radiate away \citep[e.g][]{kim2019, Lancaster2021} or spent to power outflows in other phases, such as molecular gas or X-ray hot gas. Moreover, the outflowing gas that we detect is characterized by both a high thermal energy and likely, given the large Doppler parameters $b$, a high turbulence energy too. A significant fraction of the stellar feedback energy must be used to power the thermal and turbulent motion of gas particles, which is another very important effect of feedback that prevents the gas from rapid cooling and forming stars at unrealistic high rates. 

\subsection{Spatial position of the clusters in the host galaxy}

\begin{figure}
    \centering
    \includegraphics[width=\columnwidth]{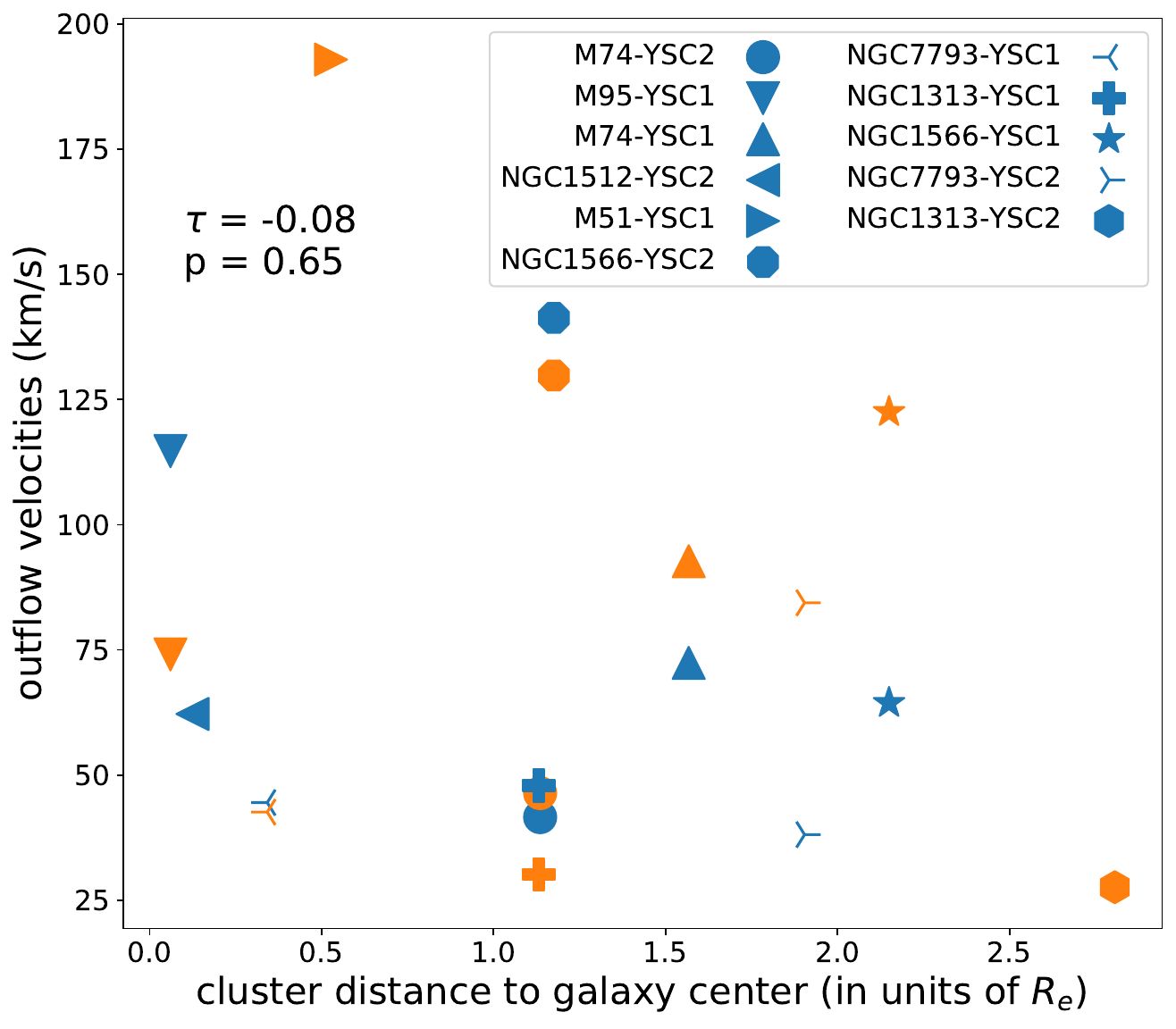}
     \caption{Outflow velocity plotted against the distance of the CLUES cluster to the center of their host galaxy in units of half-mass radii \citep{Leroy2021}. The data points are color-coded as in Figure \ref{fig:correlations} according to the phase of the gas: neutral gas in blue and ionized gas in orange. Only the spiral galaxies are included in this figure.}
    \label{fig:position}
\end{figure}

We studied the FUV absorption produced by the interstellar gas along the line of sight to YSCs, detecting 16 outflows in 18 targets. Many of the target star clusters that feature an outflow are hosted in spiral galaxies, and their location in the galaxy has been chosen to study the stellar feedback and ISM outflows as a function of the position in the galaxy. Typically, there are two target clusters per host galaxy, one cluster being close to the center of the galaxy and the other cluster being further away. In Figure \ref{fig:position} we show the outflow velocities as a function of the position of the cluster in the galaxy. This is measured calculating the distance of the star cluster to the center of the host galaxy, expressed in units of the half-mass radius, as reported by \cite{Leroy2021}. The only exception is the galaxy M51 that is not part of the sample of \cite{Leroy2021} and for which we calculated the half-light radius instead, using the available F555W HST image mosaic and finding a value of 2.9 kpc. In the assumption that the light-to-mass ratio of the galaxy is constant with the radius, the half-light radius is equivalent to the half-mass radius. 
The data points in Figure \ref{fig:position} have a large scatter and the Kendall $\tau$ analysis shows no correlation. 
In the galaxy centers, we might be seeing large scale outflows along the line of sight, which could be the case for M51-YSC1 that has the highest outflow velocity. However, the fact that the outflow kinematics correlates with the local stellar population energetics indicates that we detected small scale outflows for most of the other sources.

\subsection{Starburst-driven outflows in the literature}
\label{sec:literature_comparison}

Previous works in the literature have used FUV absorption lines to study galactic outflows, often extending to kiloparsec scales, rather than looking at individual star clusters. For the first time to our knowledge, we investigated in the UV the properties of outflowing interstellar gas at scales of $\sim$ 100 pc or less by zooming into young star clusters located in star-forming galaxies of the Local Volume (distance $<$ 16 Mpc, Paper I). 

Although we analyzed smaller physical scale outflows, we can compare column densities to galactic-scale studies, as they by definition do not depend on the area of the sky observed and therefore neither on the physical size of the object of interest. In fact, similar values of column densities of \SiII$\,$ and \SiIV$\,$ are reported by \cite{Xu2022} (Table 5), for the CLASSY survey consisting of 45 low-redshift starburst galaxies.

 Secondly, we added the averaged physical measurements of the single star clusters and their associated outflow kinematics to scaling-relations typically reported in the literature for galactic scale studies. These relations are typically used to investigate the correlation between the macro-properties of the outflows (e.g. velocities) to the rate and intensity per unit area (SFR and $\Sigma_{SFR}$) of star formation, the latter an indirect way to quantify stellar feedback and its clustering. For galactic-scale studies, the average larger distances imply that the collective feedback of several overlapping star clusters and stellar populations are responsible for driving winds and outflows. On the other hand, the CLUES outflows are driven by single star clusters and we do not expect the two samples of outflows to necessarily follow the same behavior.

\begin{figure}
    \centering
    \includegraphics[width=\columnwidth]{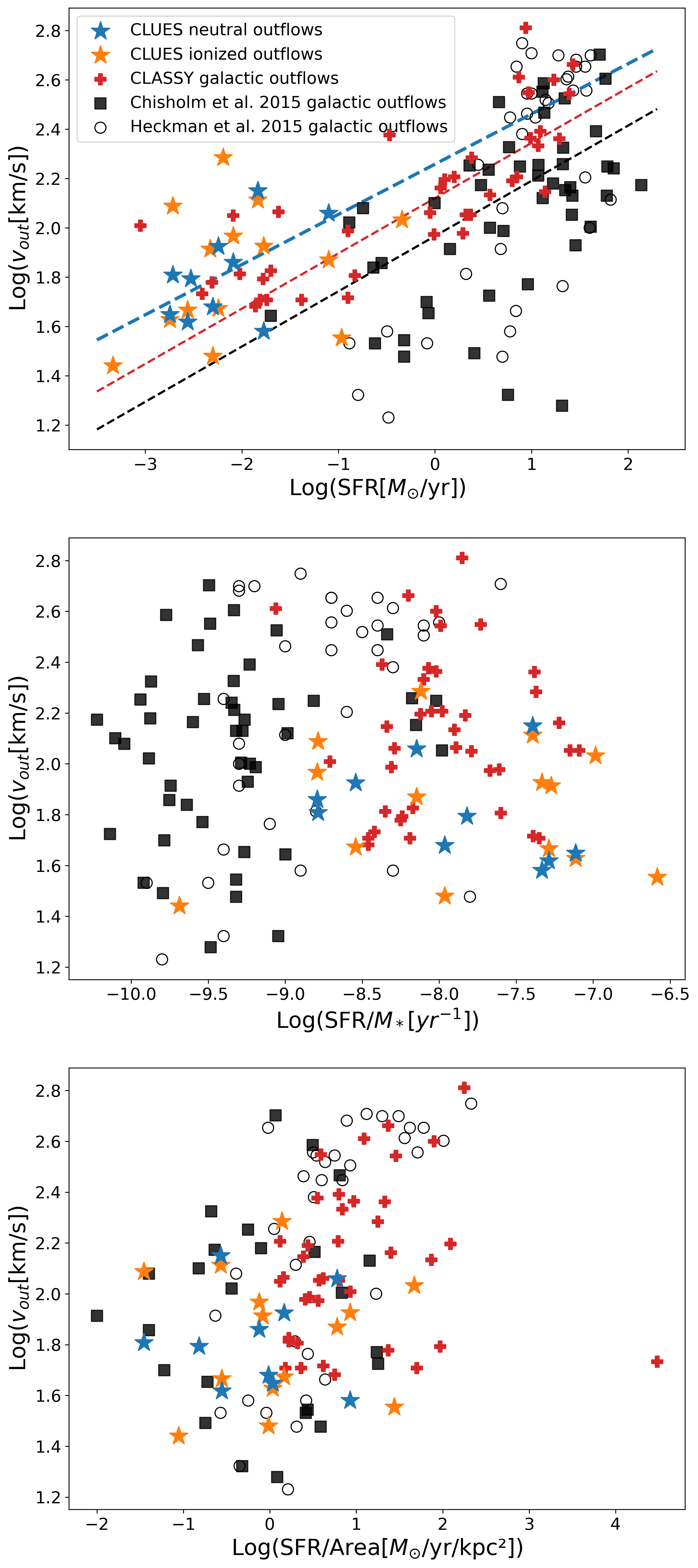}
    \caption{Outflow velocities plotted against (from top to bottom): the SFR, the SFR per unit stellar mass and the SFR per unit area. The data points plotted are taken from three different samples: in blue and orange the CLUES neutral and ionized outflows respectively, in red the CLASSY galactic outflows \citep{Xu2022}, the empty circles represent the data points of \cite{Heckman2015}, the black squares the data points of \cite{Chisholm2015}. In the top panel we report the least-square linear regression fit for the black squares points, the blue points and the red crosses, separately.}
    \label{fig:sfr}
\end{figure}

In Figure \ref{fig:sfr}, we show three different relations with data from the CLUES sample added to data available in the literature.



\cite{Heckman2015} and \cite{Chisholm2015} found a relation between the galactic outflow velocities and the SFR of the host galaxies. \cite{Xu2022} have also investigated the same relations for the CLASSY sample. For the sake of comparison\footnote{We note that the conversions from FUV Luminosity to SFR assume that SFR has been constant for $\sim$100 Myr \citep[see][]{kennicutt2012} but this is unlikely to be the case in the CLUES small apertures.}, we estimated the SFR of the CLUES clusters using the same formula as \cite{Heckman2015,Chisholm2015} based on the UV luminosity but correcting for the attenuation of the stellar populations derived in Paper I instead of using the IR luminosity. Finally, we measured the SFR per unit stellar mass and the SFR per unit area, dividing the SFR by the total stellar population mass and the COS area respectively. We note that the SFR is not well-defined for a single star cluster as the SFR is the mass of the stars formed per unit time while a star cluster is by definition a group of stars that form in a single burst (which gives an infinite value of SFR). For this reason it would be wrong to simply divide the mass of the cluster by the age. However, because the stellar population of our targets is described not only by one star cluster but also by a second component (typically an older population of diffuse stars), it makes sense to derive the SFR using the FUV luminosity as done in other studies, for the comparison with the literature numbers. 

Importantly, among the literature studies \cite{Xu2022} measured the wind velocity most similarly to us, i.e. by disentangling the stationary and the outflowing gas components. This may be the reason why CLUES outflow velocities overlap better with the \cite{Xu2022} sample rather than with \cite{Chisholm2015} and \cite{Heckman2015} samples.

Figure \ref{fig:sfr} shows the relation between the outflow velocities and the three different measurements of the star formation activity: the SFR, the SFR per unit mass and the SFR per unit area by including the CLUES sample and the literature samples described above. 
We also plotted the regression fits done by \cite{Chisholm2015} (black dashed line) and \cite{Xu2022} (red dashed line). The CLUES data points fall above these lines, meaning that when zooming-in at tens of parsec scales, we find that the star clusters launch relatively faster outflows given their low SFR compared to the galactic scale measurements. This could be a local selection effect when studying star clusters as we studied regions that are bright in the UV and active in star formation by construction. It could also be the result of the different physical scales used to estimate the SFR (with the literature studies including the whole galaxy). Overall, the CLUES data points fall close to the CLASSY outflows hosted by the lower star formation galaxies. Additionally, we plotted the regression fit to the CLUES neutral outflows (blue dashed line) with the purpose of comparing the measurements of \cite{Chisholm2015} and \cite{Xu2022}. We found a line with a slope of 2.3 similar to the slopes of 2.2 of \cite{Xu2022} and \cite{Chisholm2015} regression fits (black and red lines), but offset towards higher outflow velocities by 0.4 dex and 0.2 dex, respectively. We stress that only the CLUES neutral outflows (blues stars) and the \cite{Chisholm2015} galactic outflows (black squares) are based exclusively on the \SiII$\,$ tracer.

In the middle panel of Figure \ref{fig:sfr}, where the SFR is normalized by the total stellar mass, the CLUES data points appear to have larger values than those of \cite{Heckman2015} and \citep{Chisholm2015}. The reason for this is that while in the galactic scale measurements the whole stellar population is accounted for (including the old stars not participating in driving the outflows), for the CLUES clusters all the stars within the COS aperture are young and do power the outflows with their feedback, resulting in a high SFR per unit mass. The overlap with the CLASSY galaxies might be driven by the fact that in many of these latter galaxies the outflows are driven by a single star cluster dominating the feedback and thus comparable to our dataset.

In the lower panel, we cancelled out the physical size dependence from the star formation variable by normalizing the SFR by the area of interest. We found the CLUES data points falling in the same region of the diagram as the data points from the literature. In Table \ref{tab:kendall} we report the Kendall statistics for the three plots of Figure \ref{fig:sfr}, separately for the CLUES sample only, for the CLUES and the CLASSY samples combined, and for all the samples together. We see that the physical relations connecting SFR, SFR per unit mass and outflow velocities have similar statistical significance in the CLUES sample as those recovered for galactic scales. This however is not true for the relation between outflow velocity and SFR per unit area. The CLUES sample, even sharing a similar parameter space distribution as the galaxy samples, does not show a significant correlation.

\begin{figure}
    \centering
    \includegraphics[width=\columnwidth]{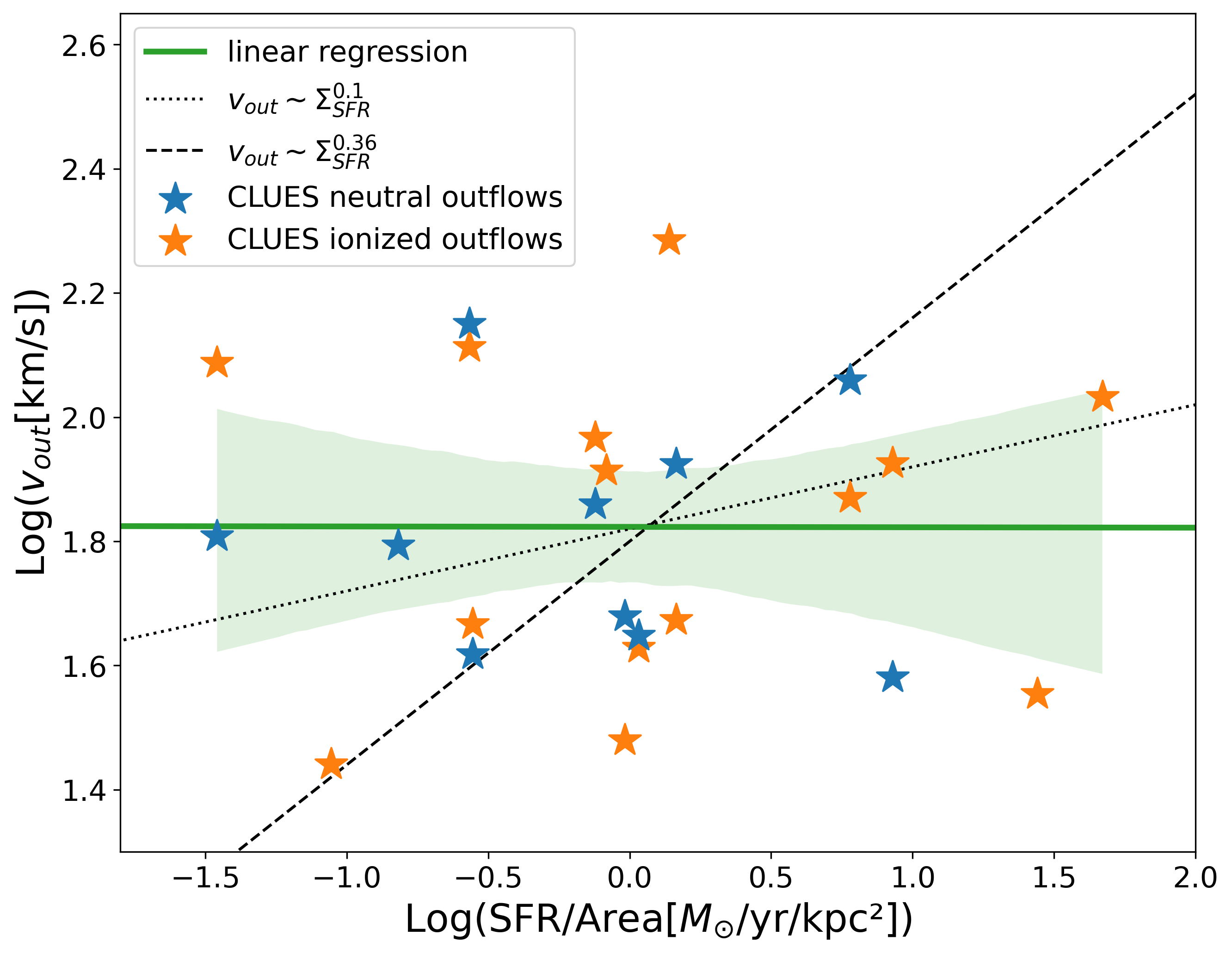}
    \caption{Linear regression of the CLUES outflow velocities plotted against SFR per unit area. The blue and orange stars mark the neutral and ionized outflows respectively, the green line is the best linear fit and light green shaded area is the 95\% confidence interval. The dotted and dashed lines are two different feedback models: energy-driven and momentum-driven, respectively.}
    \label{fig:linregr}
\end{figure}

We further investigate this latter relation by fitting a power law to the v$_{out}$ vs. $\Sigma_{SFR}$ data points of the CLUES sample. By studying this relation, our aim is to identify one of the two main feedback models \citep{Chu2022}. (i) The energy-driven feedback model predicts that v$_{out}$ scales as $\Sigma_{SFR}^{0.1}$. In this model, the outflow is primarily driven by the injection of mechanical energy from stellar winds and supernovae. (ii) The momentum-driven feedback model predicts instead a steeper relation, v$_{out}\sim\Sigma_{SFR}^{0.36}$ \citep{Zhang2018}. In the latter case, the outflow is largely powered by the radiation pressure of young stars. 
Figure \ref{fig:linregr} shows a linear regression of the CLUES outflow velocities against the SFR per unit area plotted on logarithmic scale. 
Despite the large scatter, we recover a slope of $\alpha=0.00\pm0.12$ (95\% confidence interval). This slope is more consistent with a slope $\alpha=0.1$, predicted by an energy-driven scenario \citep{Li2017,Kim2020,Li2020}, but the large scatter in the data does not allow us to reach solid conclusions. A number of integrated galaxy studies have also found shallow slopes \citep{Heckman2015,Rodriguez2019,Roberts2020} but very few works have looked at small scale outflows. Interestingly, \cite{Chu2022} found a similar result to ours for a sample of small scales outflows in a starburst galaxy studied with a resolution of 100 $\times$ 500 pc. Another similar result is the one found by  \cite{Watkins2023}, who quantified the energetics of molecular superbubbles and showed their consistency with a SN model that injects energy with a coupling efficiency of $\sim$10\%. 

In conclusion, our interpretation is that clusters look like scaled down versions of galaxies, in terms of stellar feedback and outflows. This is an interesting result, as it would imply that the kinetic effects ongoing in a galaxy are just the sum of many clusters. The consistency with energy-driven outflow models, would suggest that although radiation pressure plays an important role in the initial expansion of the bubbles (see Section \ref{sec:driving}), very likely, mechanical feedback from winds and SNe set the macroscopic physical properties we have measured at the 100s parsec scales in these bubbles. We also note that the CLUES COS sightlines might probe large scale outflows (or the sum of the cluster outflow and a larger scale outflow) which would explain the alignment of the data points in the third panel of Figure \ref{fig:sfr}. However, the correlation of outflow properties with the comparison of the cluster feedback energetics with the outflow energetics favor the cluster-scale outflow interpretation. 

\begin{table*}[]
    \centering
    \begin{tabular}{c|c|c|c}
       dataset & $v_{out}$ vs SFR & $v_{out}$ vs SFR/$M_*$ & $v_{out}$ vs SFR/Area \\
       \hline
       CLUES & $\tau$=0.30, p=0.04 & $\tau$=-0.14, p=0.36 & $\tau$=-0.03, p=0.86\\
       CLUES+CLASSY & $\tau$=0.56, p=0.000 & $\tau$=-0.05, p=0.59 & $\tau$=0.24, p=0.004\\
       All samples & $\tau$=0.44, p=0.000 & $\tau$=0.01, p=0.88 & $\tau$=0.30, p=0.000
    \end{tabular}
    \caption{Correlation coefficients (Kendall $\tau$ and p-value) for the three relations of Figure \ref{fig:sfr}: outflow velocity against SFR, SFR per unit stellar mass and SFR surface density. Three different datasets are considered: the CLUES sample (this work), the CLUES sample plus the CLASSY sample \citep{Xu2022}, and all the following samples together: CLUES, CLASSY, \cite{Chisholm2015} and \cite{Heckman2015}.}
    \label{tab:kendall}
\end{table*}

\subsection{Inflows}

Four targets in the CLUES sample, namely NGC1313-YSC2, NGC4656-YSC1, NGC4656-YSC2 and NGC5253-YSC2, feature a gas component that is inflowing towards the central star cluster with central bulk velocities of up to 75 km~s$^{-1}$. In Figure \ref{fig:voigtfit} these inflows are marked with an empty blue circle. 
We observed an inflow in those targets where the FUV dominant stellar population is oldest (30-50 Myr) and likely most of the gas in the surrounding has already been evacuated. 

Interestingly, all targets featuring an inflow show no H$\alpha$ emission (see Appendix), which is consistent with a scenario in which the star-forming region has been evacuated by a strong outflow in the past, and it is currently accreting neutral gas. However, we find puzzling that 3 of these 4 targets have both an ionized outflow and a neutral inflow. A probable explanation for this is that the observed neutral gas inflow may happen at a much larger distance to the star cluster than the ionized outflow. In such a case, the inflow is not associated with the star cluster and its stellar feedback, but rather it is part of the large scale gas kinematics of the irregular galaxy. In particular, this scenario could be valid for NGC5253-YSC2 and NGC1313-YSC2. In fact, in NGC5253 evidence was found for molecular gas in-falling towards the galaxy \citep{Meier2002} and possibly \HI$\,$ gas as well \citep{kobulnicky2008,Lopez2012}. NGC1313 instead is suggested to be interacting with a small satellite companion galaxy \citep{Silva-Villa2012}, which might perturb the large scale kinematics of the stars and gas in the galaxy.

We also notice that 3 out of 4 targets featuring an inflow are hosted by a dwarf irregular galaxy. The other target NGC1313-YSC2 is hosted by a spiral arm of a galaxy that does not have a well-defined spiral structure, and it resembles an irregular galaxy, especially in the central part. This suggests that irregular galaxies favor the occurrence of such cyclic exchange of gas with episodes of star formation followed by ISM outflows and subsequently by gas accretion via inflows. 

\section{Conclusions} \label{sec:conclusions}

In this study, we present the first FUV star cluster scale focused analyses of outflow kinematics and energetics, and investigate its impact on the interstellar gas. 

We use the stellar population models of the COS FUV spectroscopy of Paper I to constrain photon production rate ($10^{50}$ - $10^{53}$ photons/s),  mechanical energy of stellar winds and SNe combined ($10^{54}$ - $10^{55}$ erg) as well as their mechanical luminosity ($10^{39}$ - $10^{40}$ erg/s). 

We study the kinematics of the interstellar gas intervening along the line of sight to the target star clusters, by modelling the FUV absorption lines. We have employed multi-component Voigt models to fit the profile of a large variety of transitions of different ions (\AlII$\,$1670; \SiII$\,$1190,1193,1260,1526, \SII$\,$1250,1253,1259, \SiIV$\,$1393,1402, \CIV$\,$1548,1550, \NV$\,$1238,1242) with a wide range of ionization potentials (from 6.0 eV to 77.5 eV). We have carefully inspected the spectrum of each target to disentangle absorption components associated with other physical systems, such as the gas in the Milky Way or in the Magellanic System. Four different \SiII$\,$ transitions of different oscillator strengths have been used to disentangle column density and covering fraction, which both contribute to the absorption of the continuum source flux. We find gas covering fractions close to 1 for all targets, ruling out hidden saturation as a potential issue. 
We derive velocities, column densities and $b$ parameters for each component of low and high ionization transitions of several ions. We detect 16 outflows in 18 targets, and in the majority of the cases the outflows are both in the neutral phase (traced by \SiII) and ionized phase (traced by \SiIV). Outflow velocities span a range between 30 and 190 km~s$^{-1}$. We detect a gas component in the neutral phase that is inflowing towards the young star cluster in 4 out of 18 targets (NGC1313-YSC2, NGC4656-YCS1, NGC4656-YSC2 and NGC5253-YSC2). For all these targets except NGC4656-YSC1, we detect an ionized outflow too, which suggests that the inflow is not associated with the star cluster but is tracing a larger scale gas kinematics.

We investigate the presence of correlations between outflow velocities and stellar properties. We find significant correlations between the neutral gas outflow velocities and the following stellar properties: mass of the stellar populations, photon production rate and mechanical luminosity of the stellar feedback. These correlations are expected and consistent with each other, since a more massive young stellar population means more ionizing photons and mechanical energy produced per unit time. We do not find correlations with age and mechanical energy of the stellar feedback from the total stellar population. We note that outflow velocities are expected to depend on many parameters at once (cluster age, cluster mass, feedback momentum and energy injected, ISM density), and the sample is not large enough to hold some of the parameters fixed while exploring others.
 Only a weak correlation between ionized gas outflow velocities and the mechanical luminosity of stellar winds is observed. The absence of strong correlations for the ionized gas outflows is in part due to the presence of the outlier M51-YSC1, but likely also due to the larger uncertainties of the outflow measurements that result in a larger scatter of the data points. 
The correlation of the neutral outflow velocities with the photon production rate and the mechanical luminosity of stellar winds indicate the importance of pre-SN feedback in re-processing the gas of the natal cloud where SN will explode \citep{Lopez2014,McLeod2021,Barnes2021,Chevance2022,DellaBruna2022,Bending2022}.

We compare the energetics of the outflows with those estimated from the stellar populations, albeit with large uncertainties due to the assumptions of spherical symmetry and uniform density with the errors on the outflow radii. The ratio between the kinetic power of the CLUES outflows and the mechanical luminosity of the stellar feedback is 10$\pm$8 \%. This is consistent with what found in galactic scale outflow studies \citep[1-20 \%][]{Chisholm2017}, and theoretical studies \citep[1-10 \%][]{Lancaster2021b}, as well as what reported from star-forming regions in local spiral galaxies \citep[8-50 \%][]{Chevance2022}.

We compare the CLUES outflows at sub-kpc scales, driven by star clusters, with the galactic outflows studied by \cite{Heckman2015}, \cite{Chisholm2015} and \cite{Xu2022} in an SFR vs outflow velocity diagram. In the diagram of SFR per unit area vs outflow velocity, the CLUES data points lie in the same region as the data points from the literature, suggesting a universal powering mechanism independent of scales for launching gas outflows. We do not find any statistically significant correlation between SFR per unit area vs outflow velocity when only the CLUES data points are analyzed. We notice that a very shallow correlation is predicted in case of an energy-driven outflow model \citep{Li2017,Kim2020,Li2020}, where mechanical energy from winds and SN are the major drivers. Our data are consistent with this shallow trend, but the large scatter does not allow us to reach any definitive conclusion. Combined with the observed correlation between outflow velocity and pre-SN feedback, our study points toward the importance of both the photoionization and the mechanical feedback in setting the observed macroscopic properties of these 100s pc scale outflows.

Our work provides further observational constraints on the physics of outflows, importantly at the small scales where feedback originates. In future works, bigger samples will be needed to tease apart the effects of cluster age, cluster mass, metallicity and ISM density.

\newpage

\section{Acknowledgments}.
\begin{acknowledgments}
This research is based on observations made with the NASA/ESA Hubble Space Telescope obtained from the Space Telescope Science Institute, which is operated by the Association of Universities for Research in Astronomy, Inc., under NASA contract NAS 5–26555. These observations are associated with program(s) 15627, 11579.
This research has made use of the NASA/IPAC Extragalactic Database (NED), which is funded by the National Aeronautics and Space Administration and operated by the California Institute of Technology.
We thank Jens-Kristian Krogager for the support with using his software VoigtFit and adapting it to our purpose.  We thank David French for sharing the Python code to make Figure 2. A.A and M.S. acknowledge support from the Swedish National Space Agency (SNSA) through the grant Dnr158/19. M.H. is fellow of the Knut \& Alice Wallenberg Foundation. K.G. is supported by the Australian Research Council through the Discovery Early Career Researcher Award (DECRA) Fellowship (project number DE220100766) funded by the Australian Government. 
K.G. is supported by the Australian Research Council Centre of Excellence for All Sky Astrophysics in 3 Dimensions (ASTRO~3D), through project number CE170100013. 
\end{acknowledgments}

%

\vspace{5mm}
\facilities{HST(COS), VLT(MUSE).}


\software{astropy \citep{2013A&A...558A..33A,2018AJ....156..123A},
          VoigtFit \citep{Krogager2018}
}



\appendix
\label{app:figures}

The figures in this appendix show the kinematics of the interstellar gas along the line of sight to the CLUES star clusters. All targets are reported in the appendix figures except M95-YSC1 and NGC4656-YSC1 that are reported in Figure \ref{fig:example_outflow_inflow}.

\begin{figure}
    \includegraphics[width=\columnwidth]{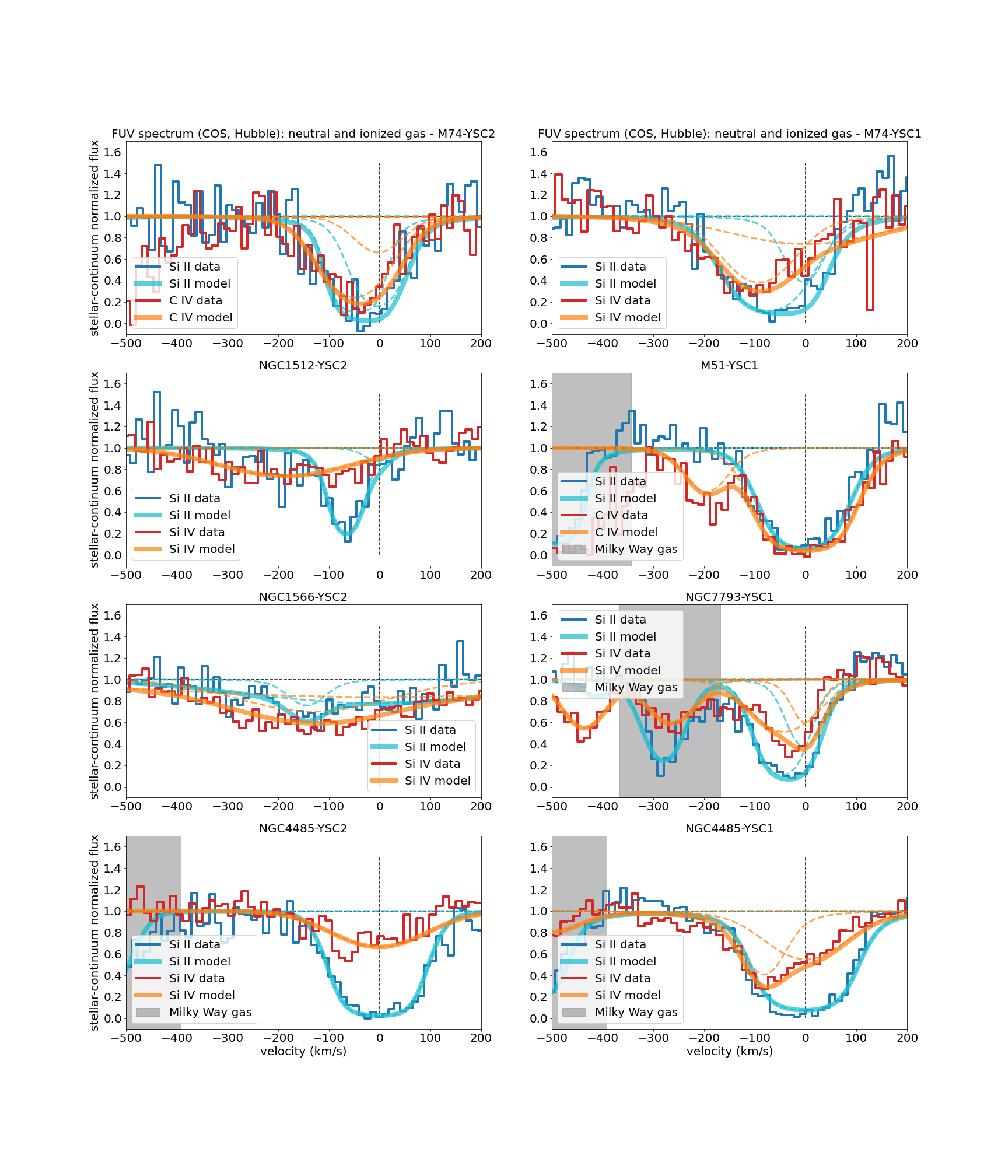}
    \label{fig:a1}
    \caption{See caption of Figure \ref{fig:a2} (continuation).}
\end{figure}

\begin{figure}
    \includegraphics[width=\columnwidth]{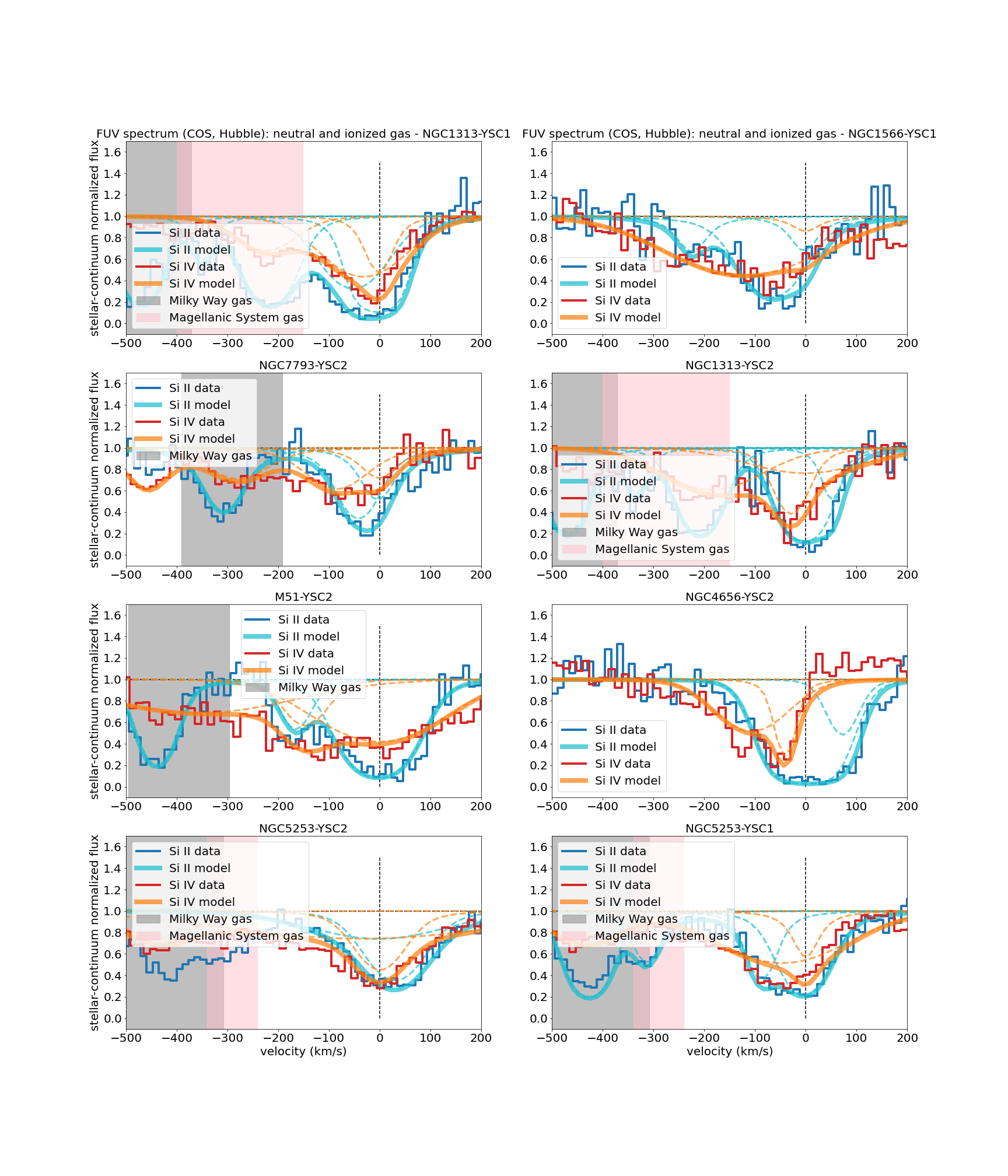}
    \label{fig:a2}
    \caption{Kinematics of neutral and ionized gas surrounding the CLUES clusters. \SiII$\,$ spectrum is shown in blue and \SiIV$\,$ (or \CIV) is shown in red/orange, including data, model and model components (dashed lines). Regions of the spectrum with absorption contamination from the Milky Way or Magellanic System gas are marked with grey and pink shaded areas, respectively.}
\end{figure}


\newpage

\bibliography{sample631}{}
\bibliographystyle{aasjournal}



\end{document}